\documentclass[10pt,aps,pre,amsmath,amsfonts,amssymb,floatfix,longbibliography,superscriptaddress,notitlepage, twocolumn]{revtex4-1}
\usepackage{mathrsfs} \usepackage{graphicx,color} \usepackage{bbm} \usepackage{multirow}
\usepackage[normalem]{ulem}
\usepackage{lineno}
\usepackage{tabularx}

  \def\erf{\text{erf}}

\newcommand{\redd}[1]{\textcolor{black}{#1}}
\newcommand{\red}[1]{\textcolor{black}{#1}}

\def\to{\rightarrow}

\newcommand{\eq}[1]{Eq.~(\ref{#1})}
\newcommand{\beq}{\begin{equation}} \newcommand{\eeq}{\end{equation}}
\newcommand{\bmat}{\begin{displaymath}} \newcommand{\emat}{\end{displaymath}}

\renewcommand{\vec}[1]{\boldsymbol{\mathrm{#1}}}
\def\vr{{\boldsymbol r}}

\begin{document}

\title{A jamming plane of sphere packings}




\author{Yuliang Jin}
\email{yuliangjin@mail.itp.ac.cn}
\affiliation{CAS Key Laboratory of Theoretical Physics, Institute of Theoretical Physics, Chinese Academy of Sciences, Beijing 100190, China}
\affiliation{School of Physical Sciences, University of Chinese Academy of Sciences, Beijing 100049, China}
\affiliation{Cybermedia Center, Osaka University, Toyonaka, Osaka 560-0043, Japan}

\author{Hajime Yoshino}
\email{yoshino@cmc.osaka-u.ac.jp}
\affiliation{Cybermedia Center, Osaka University, Toyonaka, Osaka 560-0043, Japan}
\affiliation{Graduate School of Science, Osaka University, Toyonaka, Osaka 560-0043, Japan}

\begin{abstract}
  The concept of jamming has attracted great research interest due to its broad relevance in soft-matter such as liquids, glasses, colloids, foams, and granular materials, and its deep connection to {sphere packing  and
optimization problems.}
  Here we show that the
domain of 
 amorphous jammed states 
 of frictionless spheres 
 can be significantly extended, from the well-known jamming-point at a fixed density,
 to a {\it jamming-plane} that spans
 the density and shear strain axes.
We explore  the jamming-plane, via athermal and thermal simulations of  compression and shear jamming,
{with initial equilibrium configurations prepared by an efficient swap algorithm.}
The jamming-plane can be divided into reversible-jamming and irreversible-jamming regimes, based on the reversibility of the route from the initial configuration to jamming.
{Our results suggest that}
the irreversible-jamming behavior 
reflects an escape from the meta-stable glass basin to which the 
  initial configuration belongs to, or the absence of such
   basins.
   All jammed states, either compression or shear jammed, 
   are isostatic, and exhibit jamming criticality of the same universality class.
However,  the anisotropy of contact networks non-trivially depends on the jamming density and  strain.
Among all state points on the jamming-plane, the jamming-point
is a unique one with
the minimum jamming density and the maximum randomness. 
{For \red{crystalline}
  packings}, 
the jamming-plane shrinks into a single shear jamming-line that is independent of  initial configurations.
 Our study paves the way for solving the long-standing random close packing problem, and provides a more complete framework to understand jamming.
\end{abstract}

\maketitle


In three dimensions, 
{the densest packing of equal-sized spheres has a face-centered cubic (FCC) or a hexagonal close packing (HCP) structure, and the density  (packing fraction)  is $\varphi_{\rm FCC} = \varphi_{\rm HCP} \simeq 0.74$.}
This was conjectured initially by the celebrated scientist 
Kepler in the 17th-century, known as the Kepler conjecture, and was proved by mathematician 
Hales about 400 years later~\redd{\cite{hales2005proof}}.

The ``random version'' of the sphere packing problem, however, remains unsolved. 
In the 1960th, based on the empirical observation that the packing fraction of ball bearings, when poured, shaken, or kneaded inside balloons,
never exceeds  a maximum value $\varphi_{\rm RCP} \approx 0.64$, Bernal introduced the concept of {\it random close packing} (RCP) to characterize the optimal way to pack spheres randomly~\cite{bernal1960packing}. Although many experiments and simulations have reproduced random packings with a volume fraction around 0.64, an agreement on the exact value of $\varphi_{\rm RCP}$ has not been reached. Torquato et al. proposed that the idea of RCP should be replaced by a new notion called {\it maximally random jammed} (MRJ) state, where the randomness is measured by some order parameters which characterize the crystalline order~\cite{torquato2000random}. O'Hern  et al. designed a fast quench protocol {that} generates randomly jammed {isotropic} packings of monodisperse spheres  at $\varphi_{\rm J} = 0.639 \pm 0.001$ in the thermodynamic limit~\cite{o2003jamming}, which is 
referred to as the jamming-point (J-point)~\cite{liu1998nonlinear}. Later, based on mean-field calculations, Zamponi and Parisi predicted that the jamming density of amorphous packings should span
over a range on the 
 jamming-line (J-line)~\cite{parisi2010mean, parisi2020theory}, which has been supported in a number of numerical simulations~\cite{speedy1996distribution, chaudhuri2010jamming, ozawa2012jamming, ozawa2017exploring}.

{Spheres can be 
constrained not only by compression but also by shear.
\redd{Many   experiments and simulations have reported shear jamming in  granular matter and suspensions  ~\cite{cates1998jamming,bi2011jamming, vinutha2016disentangling, seto2019shear, kumar2016memory, das2020unified, zhao2019shear, babu2020friction,kawasaki2020shear}, and  it was suggested by several studies that frictional interactions are essential for shear jamming~\cite{bertrand2016protocol, baity2017emergent} .
Different from compression jamming, the contact network of a shear jammed packing is generally anisotropic.}


\begin{figure}[ht]
\centerline{\includegraphics[width=\columnwidth]{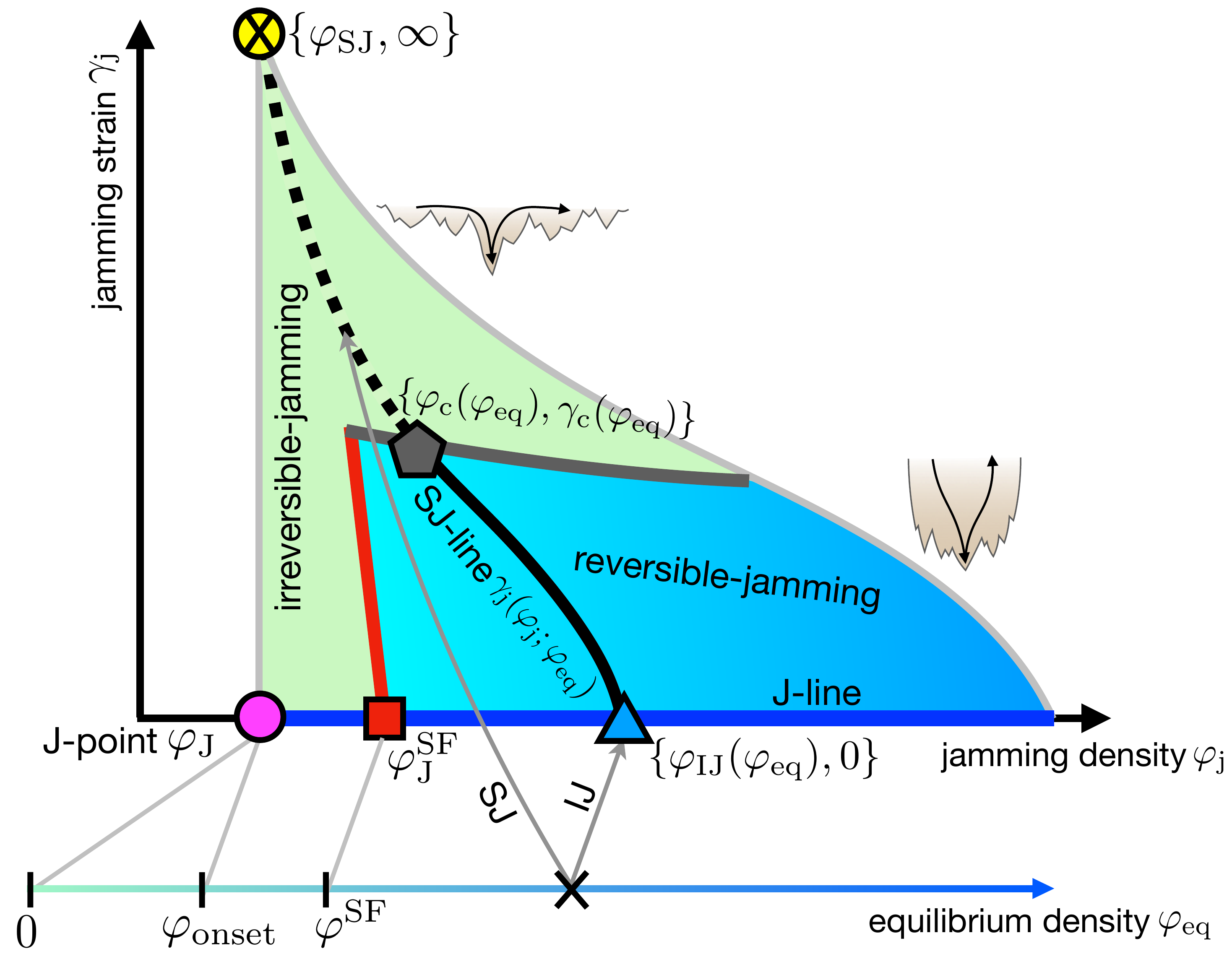}} 
\caption{{Schematic J-plane.}  The J-plane
  {$(\varphi_{\rm j},\gamma_{\rm j})$}
is {the region where  frictionless amorphous jammed {configurations}
  at {jamming} density $\varphi_{\rm j}$ and  {jamming} strain $\gamma_{\rm j}$ {exist}.} The J-point and J-line known from the previous studies
are included as its subset.
  \red{Isotropic jamming (IJ) and shear jamming (SJ) protocols  bring the system, which is initially
   in  a thermalized HS glass state at
    a given
      density $\varphi_{\rm eq}$ (an example is indicated by the cross mark on the $\varphi_{\rm eq}$-axis),
to jammed states along the associated SJ-line $\gamma_{\rm j}(\varphi_{\rm j}; \varphi_{\rm eq})$, whose end points are $\{\varphi_{\rm IJ}(\varphi_{\rm eq}), 0 \}$ and $\{ \varphi_{\rm SJ} = \varphi_{\rm J} , \infty \}$.}
The J-plane is divided into reversible-jamming and irreversible-jamming  regimes. {Reversible and irreversible jamming routes are demonstrated schematically on  corresponding free-energy landscapes.}
\red{The arrows indicate  how the system evolves  on the free-energy landscape in a cycle of compression/shear (unjammed $\to$ jammed $\to$ unjammed). The jamming routine is reversible if the system remains in the same meta-stable glass basin after the cycle.}
Each state in the reversible-jamming regime
can be related reversibly to the initially thermalized
HS liquid state at $(\varphi_{\rm eq},0)$. The lighter color in this region represents the smaller corresponding $\varphi_{\rm eq}$.
The reversible-jamming regime is
upper-bounded by the yielding-jamming separation line {of thermal HSs} $\{\varphi_{\rm c}(\varphi_{\rm eq}), \gamma_{\rm c}(\varphi_{\rm eq})\}$ (gray line), and left-bounded by the state-following line ({red} line), whose end point is the  state-following  jamming point at $\{\varphi_{\rm J}^{\rm SF}, 0 \}$.
{How the J-line and the J-plane are bounded from above is an open question.
\red{The relationship between the isotropic jamming density $\varphi_{\rm IJ}$ and the initial equilibrium density $\varphi_{\rm eq}$ (see Figs.~S5 and~S9) are visualized by the connections between the $\varphi_{\rm eq}$-axis and the $\varphi_{\rm j}$-axis; in particular, $\varphi_{\rm IJ} (0  \leq \varphi_{\rm eq} \leq  \varphi_{\rm onset}) =\varphi_{\rm J}$ and  $\varphi_{\rm IJ}( \varphi_{\rm eq} = \varphi^{\rm SF}) = \varphi_{\rm J}^{\rm SF}$ (see  Table~\ref{table:densities} for the definitions and values of these characteristic densities).}  
}
}
\label{fig:Jplane_sche}
\end{figure}

\red{In this study, we propose, and construct through numerical simulations,  a  jamming-plane (J-plane), which extends the domain of \redd{fritionless} jammed states form the previously established J-point~\cite{o2003jamming} and J-line~\cite{parisi2010mean}, by including anisotropic states realized by shear jamming. The J-plane provides a framework to clarify the answers to the following two important questions (i) and (ii).}


{\red{(i) {\bf Do isotropic and shear jamming occur at the same density in frictionless spheres, in the thermodynamic limit?} Previously, several simulation}  studies~\cite{bertrand2016protocol, baity2017emergent} showed that, for an infinitely large, unjammed, frictionless sphere system at  a density $\varphi$ below the isotropic jamming density  $\varphi_{\rm J}$, the probability to obtain jamming is zero at any applied shear strain, which leads to a conclusion that shear jamming \red{necessarily occurs at the unique density $\varphi_{\rm J}$ in the thermodynamic limit.}
 In particular, the conclusion was based on finite-size analyses  of data obtained from athermal quasi-static shear (AQS) simulations, where random configurations were used as initial conditions~\cite{bertrand2016protocol, baity2017emergent}.
On the other hand,
  a recent mean-field theory~\cite{urbani2017shear}
  predicts the existence of shear jamming \red{at different densities} in deeply annealed, frictionless, thermal hard sphere (HS) glasses.}  

To clarify the issue, in this paper 
we systematically examine the possibility of shear jamming of frictionless spheres,  in strain-controlled~\cite{vinutha2016disentangling, bertrand2016protocol, baity2017emergent} and stress-controlled~\cite{babu2020friction} AQS simulations of soft spheres (SSs), as well as in strain-controlled thermal quasi-static shear (TQS) simulations of HSs~\cite{jin2017exploring, jin2018stability}. Inspired by~\cite{urbani2017shear}, particular attention is paid
to the impact of
  the glass transition
  on the routes to jamming~\cite{parisi2010mean}. 
As a result, we show that frictionless shear jamming exists \red{over a range of densities},
when deeply supercooled liquid  configurations (with crystallization suppressed),
instead of purely random (or poorly annealed) configurations as in~\cite{bertrand2016protocol, baity2017emergent}, are used as the initial conditions.
This is confirmed by 
all kinds of jamming protocols performed in this study, with finite-size analyses.
We find that frictionless shear jamming
could take place in
unjammed sphere assemblies {\it denser} than the J-point density $\varphi_{\rm J}$, which is the minimum possible jamming density.
It means that, if the shear is reversed, these systems will firstly unjam and then
jam again but in the opposite direction, much as 
in the frictional case \cite{cates1998jamming}.
%

\red{It should be noted that deeply annealed glasses are accessible by multiple protocols.}
Besides the thermal annealing approach used here and previously in~\cite{chaudhuri2010jamming, ozawa2012jamming, ozawa2017exploring}, 
unjammed configurations above $\varphi_{\rm J}$ can 
be also obtained by protocols like athermal cyclic over-compression~\cite{kumar2016memory} or athermal cyclic shear~\cite{das2020unified, babu2020friction}. 
These mechanical training processes can be considered as effective annealing~\cite{babu2020friction}, and are reproducible in experiments.
For example, it is possible to implement AQS using the newly developed 
multi-ring Couette shear set-up~\cite{zhao2019shear}.

\red{(ii) {\bf 
    Static jamming can take place at various densities depending on compression protocols~\cite{speedy1996distribution, chaudhuri2010jamming, ozawa2012jamming, ozawa2017exploring, parisi2010mean, parisi2020theory}, while the dynamic jamming density, or the {\it critical state density} $\varphi_{\rm c}$,  obtained  in stationary shear rheology, was shown to be unique~\cite{olsson2007critical, hatano2008scaling, otsuki2012rheology, rahbari2018characterizing} --
    how can the two seemingly paradoxical observations be reconciled?}}

  \red{Under compression, the jamming density depends on the initial condition -- the deeper the initial state is annealed, the higher jammed density is obtained. This ``memory effect'' is examined by detailed reversibility tests in this paper. It turns out that the states at $\varphi_{\rm J}$ are irreversibly jammed and therefore {\it memoryless}.}

\red{On the other hand , under shear, \red{even} a deeply annealed system \red{can be} rejuvenated and eventually evolves into a steady state where the initial memory is completely lost. In the quasi-static and thermodynamic limits, the viscosity of steady states diverges at $\varphi_{\rm c}$ from below jamming, and the yield stress vanishes at $\varphi_{\rm c}$ from above. It was shown in Ref.~\cite{babu2020friction} that, the same equations of states are shared by the steady states under shear and the isotropically jammed states compressed from $\varphi_{\rm J}$, and the two jamming densities are very close to each other, $\varphi_{\rm c} \simeq \varphi_{\rm J}$.}

\red{Here, after exploring both compression and shear jammed states in a systematic and well-controlled way, we find that, the density $\varphi_{\rm c} \simeq \varphi_{\rm J}$ of memoryless (under both compression and shear) jammed states sets a lower density bound for frictionless jamming. Higher jamming densities are  obtained only if the initial memory can be kept;  the initial condition (degree of annealing) is irrelevant anymore for memoryless states at $\varphi_{\rm c} \simeq \varphi_{\rm J}$. }

 \red{The properties of  new states, obtained by shear jamming, 
 deserve to be analyzed in detail.}
\red{For examples, is} shear jamming
 reversible upon {reverting} the  route to jamming~\cite{jin2018stability}? 
 Are shear jammed packings isostatic {(i.e., the average contact number per particle is $z_{\rm j}=2d = 6$ in $d=3$ dimensions}) as in the isotropic jamming case ~\cite{o2003jamming, torquato2010jammed}?
 Do {compression and shear jammed packings}
{exhibit critical properties of}
 the same jamming universality class~\cite{charbonneau2014fractal}?
How do the anisotropy~\cite{radjai1998bimodal, bi2011jamming} and the bond-orientational order~\cite{steinhardt1983bond} of the contact network change with the jamming strain? What is the difference on shear jamming between amorphous states and crystals? All these questions will be answered in this paper.\\

{\bf \Large Results}\\

{\bf  Construction of the  Jamming-plane}

{We simulate a thermal HS model and an athermal SS model, which have the same continuous diameter distribution but different inter-particle interactions (see Supplementary Information (SI) Sec.~S1). 
Typically, a critical jammed packing can be considered as a HS system with zero nearest-neighbor separations, or equivalently a SS system with zero nearest-neighbor overlappings.
Choosing these models has a major advantage:
by using an efficient swap algorithm~\cite{BCNO2016PRL}, it is possible to prepare equilibrium (liquid) HS configurations over an extremely wide range of equilibrium densities $\varphi_{\rm eq}$ (see SI Sec.~S2). 
The timescale of the most deeply annealed states (largest  $\varphi_{\rm eq}$)  significantly exceeds previous limitations in simulations,  becoming even comparable to experimental scales~\cite{berthier2017configurational}. This breakthrough opens a whole new realm of possibilities for exploring physics in previously inaccessible domains~\cite{BCNO2016PRL, berthier2016growing, berthier2017configurational, ninarello2017models, jin2017exploring, jin2018stability, babu2020friction}, including isotropic jamming~\cite{coslovich2017exploring}. 
In this study, we apply the same methodology to study shear jamming, showing that  such a quantitative advance can conceptually change our understanding: shear jamming is in fact possible in {\it frictionless} spheres, if deeply supercooled liquid
configurations (more specifically, when $\varphi_{\rm eq}$ is above a certain threshold $\varphi_{\rm onset}$) are used as the initial conditions in the jamming protocols.}

{Without loss of generality, one can assume that any sphere packing is jammed at a {\it jamming density} $\varphi_{\rm j}$ and a {\it jamming strain} $\gamma_{\rm j}$, with isotropic jamming being the special cases when $\gamma_{\rm j} = 0$. Therefore, a strain-density ($\gamma_{\rm j}-\varphi_{\rm j}$) J-plane, as demonstrated schematically in Fig.~\ref{fig:Jplane_sche}, provides a  parameter space that can include all packings jammed by either compression or shear protocols. 
We emphasize that, in this study, $\varphi_{\rm j}$ is used to represent the jamming density of any packings, while  $\varphi_{\rm J}$ is the unique  J-point density, which in general satisfies $\varphi_{\rm J}\leq \varphi_{\rm j}$.
To map out the J-plane numerically (Fig.~\ref{fig:Jplane_comparison}a), we employ four standard jamming protocols (SI Sec.~S3): athermal rapid compression (ARC)~\cite{o2003jamming}, strain-controlled AQS~\cite{bertrand2016protocol, baity2017emergent}, thermal compression (TC)~\cite{lubachevsky1990geometric, skoge2006packing, donev2005pair} and strain-controlled TQS~\cite{jin2018stability}. A further confirmation is  obtained by stress-controlled AQS simulations (SI Sec.~S7). Below we briefly describe main features of the J-plane.}\\



\begin{figure}[ht]
\centerline{\includegraphics[width=1.0\columnwidth]{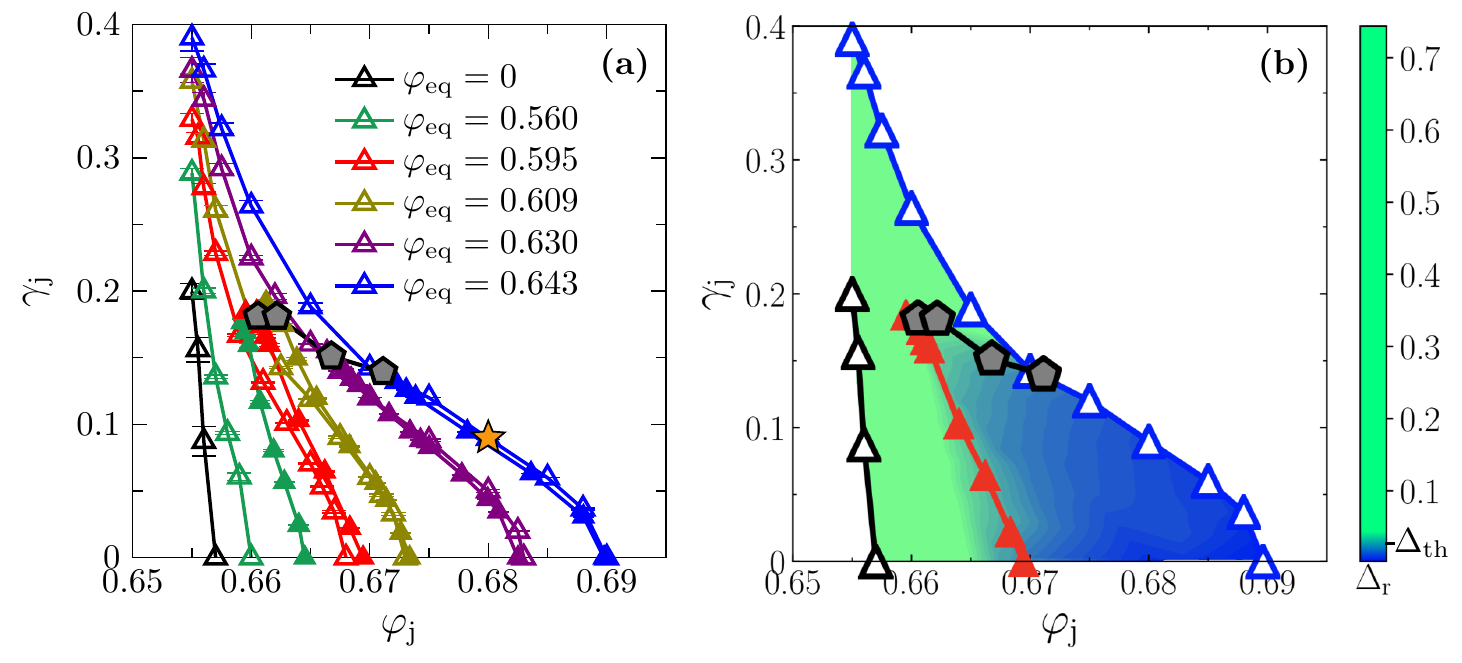}} 
\caption{ 
{Numerical J-plane.}
(a) Numerical data of the SJ-lines, for $N=8000$ {spheres} and  a few different  $\varphi_{\rm eq}$ (same data as in SI Figs.~S7 and~S12).  The filled 
and open symbols correspond to {data obtained from} thermal and athermal protocols, respectively.
{The star represents the state point $\{ \varphi_{\rm j} = 0.68, \gamma_{\rm j} = 0.09 \}$ (for $\varphi_{\rm eq} = 0.643$) examined by the stress-controlled protocol in SI Fig.~S15.}
 (b) J-plane colored according to the {RMSD} $\Delta_{\rm r}$ measured by one cycle of AQS  \red{(see Fig.~S13 for the detailed data of $\Delta_{\rm r}$)}. {The route to jamming is reversible/irreversible if $\Delta_{\rm r}$  is below/above a threshold $\Delta_{\rm th} =  0.025$}.
 The reversible-jamming (blue) and the irreversible-jamming (green) regimes are separated by the yielding-jamming separation line (pentagon line), and the state-following  line corresponding to the thermal SJ-line for $\varphi_{\rm eq} = \varphi_{\rm SF}$ (red filled triangle line). {Data are averaged over 100-2000 independent samples. Error bars represent the standard error of the mean in this paper.}}
\label{fig:Jplane_comparison}
\end{figure}


{The J-plane is a collection of all state points $\{\varphi_{\rm j}, \gamma_{\rm j} \}$  representing amorphous jammed packings.
Our simulation results are summarized in the numerical J-plane Fig.~\ref{fig:Jplane_comparison}(a), 
and in SI we explain in detail how they are obtained by using athermal (see SI Sec.~S4)  and thermal (see SI Sec.~S5) jamming protocols. 
Note that,  the J-plane may quantitatively depend on model parameters such as the polydispersity, and it does not include partially ordered packings~\cite{torquato2000random, jin2010first}, as crystallization is highly suppressed by the large polydispersity in our models~\cite{BCNO2016PRL}. The J-plane contains the following three elements.}

(i) {Jamming-point}. 
The J-point~\cite{o2003jamming} at $\{ \varphi_{\rm J}, 0 \}$  
{is a special, unique point on the J-plane, because its density $\varphi_{\rm J}$ is the lowest possible density among all state points.}  
The packings at the J-point can be  generated by {ARC} from random initial configurations
  {with $\varphi_{\rm eq}=0$},  {as done in~\cite{o2003jamming} (see SI Sec.~S4A), or more generally for any initial configurations with $\varphi_{\rm eq} \leq \varphi_{\rm onset}$~\cite{ozawa2012jamming}, where $\varphi_{\rm onset} \approx 0.56$ is the onset density of glassy dynamics~\cite{berthier2017configurational} (see SI Fig.~S5).
  For our model, the finite-size analysis (SI Figs.~S3 and~S4) gives $\varphi_{\rm J} = 0.655(1)$ in the thermodynamic limit (see Table~\ref{table:densities} for a summary of  all relevant densities). }
  The  vertical line $\{\varphi_{\rm j} = \varphi_{\rm J}, \gamma_{\rm j} \}$
{(the gray vertical line starting from the J-point in Fig.~\ref{fig:Jplane_sche})} sets the leftmost boundary of the J-plane: no packing exists below $\varphi_{\rm J}$ {in the thermodynamic limit}. {The packings shear jammed below $\varphi_{\rm J}$ are due to the finite-size effect (see SI Fig.~S3), as previously noticed in ~\cite{bertrand2016protocol, baity2017emergent}. } 

  (ii) {Jamming-line}. The  J-line $\{\varphi_{\rm IJ},0\}$
  {(the blue line at the bottom of Fig.~\ref{fig:Jplane_sche})} with
\beq
  \varphi_{\rm IJ}=\varphi_{\rm IJ}(\varphi_{\rm eq}),
  \label{eq:Jline}
  \eeq
is formed by the state points of isotropic packings. {The isotropic packings are obtained by isotropic compression protocols (ARC or TC), and the isotropic jamming density $\varphi_{\rm IJ}$
depends on the density $\varphi_{\rm eq}$ of the initial equilibrium HS configuration before compression (see SI Sec.~S4B and Sec.~S5C 
for how to obtain the J-line numerically)
~\cite{chaudhuri2010jamming, ozawa2012jamming, ozawa2017exploring}.}

{While the lower bound of J-line, $\varphi_{\rm IJ} = \varphi_{\rm J}$ can be obtained by ARC,  the minimum isotropic jamming density generated  by TC is $\varphi_{\rm IJ}^{\rm min, th} = 0.665$, which is above $\varphi_{\rm J}$ due to thermal activations (see SI Sec.~S5A).} 

According to the mean-field theory~\cite{parisi2010mean}, the J-line \eq{eq:Jline} is bounded from above by 
 the {\it glass close packing} density, $\varphi_{\rm GCP} = \varphi_{\rm IJ}(\varphi_{\rm K})$, where $\varphi_{\rm K}$ is the Kauzmann point density. However, reaching this point is beyond {the current computational power}.
 In practice, the maximum {isotropic} jamming density $\varphi_{\rm IJ}^{\rm max}$ depends on the protocol efficiency. 
 For example, athermal training protocols  typically give  $\varphi_{\rm IJ}^{\rm max} \approx \varphi_{\rm J} + 0.02$~\cite{kumar2016memory, das2020unified}.
In this study,  we are able to reach 
 {$\varphi_{\rm IJ}^{\rm max} \approx \varphi_{\rm J} + 0.035$, thanks to the efficient swap algorithm}~\cite{ozawa2017exploring}. 

For clarity, let us note again that there are other protocols
  to prepare initial configurations, equilibrated to certain extents,
  such as the cyclic compression \cite{kumar2016memory} and the cyclic shear
 protocols  \cite{das2020unified}.
For them,
$\varphi_{\rm IJ}$ depends on parameters that are specific to the protocol.
{For example,}
in the cyclic compression protocol, $\varphi_{\rm IJ}(\varphi^{\rm max})$ depends on the maximum over-compression density $\varphi^{\rm max}$~\cite{kumar2016memory}, {and} in the cyclic shear protocol, $\varphi_{\rm IJ}(\gamma^{\rm max})$ depends on the maximum shear strain $\gamma^{\rm max}$~{~\cite{das2020unified}}.

(iii)  {Shear jamming-lines (SJ-lines)}. 
Each SJ-line {(e.g., the curved line connecting the yellow circle
  and the blue triangle in Fig.~\ref{fig:Jplane_sche})},
  \beq
  \gamma_{\rm j}=\gamma_{\rm j}(\varphi_{\rm j}; \varphi_{\rm eq}),
  \label{eq:SJline_1}
\eeq
or equivalently,
  \beq
  \varphi_{\rm j}=\varphi_{\rm j}(\gamma_{\rm j}; \varphi_{\rm eq}),
    \label{eq:SJline_2}
\eeq
represents the functional dependency between $\gamma_{\rm j}$ and $\varphi_{\rm j}$  
 for a given $\varphi_{\rm eq}$.
 The states 
 with a non-zero jamming strain $\gamma_{\rm j}$ are
 {said to be {\it shear jammed}}. 
The lower end point of the SJ-line at $\gamma_{\rm j}=0$ is nothing but an 
isotropic jamming point
at $\{\varphi_{\rm IJ}(\varphi_{\rm eq}), 0 \}$
on the J-line \eq{eq:Jline}, {which is identical to,}
\beq
\varphi_{\rm IJ}(\varphi_{\rm eq})=\varphi_{\rm j}(0; \varphi_{\rm eq}),
\eeq
and the upper  end point is at $\{\varphi_{\rm SJ} = \varphi_{\rm J}, \gamma_{\rm j} = \infty \}$, {in the thermodynamic limit (see SI Fig.~S6),} where all SJ-lines meet.  
The J-plane contains infinite number of SJ-lines,
but {numerically we only use} a few typical SJ-lines
to represent the J-plane {as shown in Fig.~\ref{fig:Jplane_comparison}(a)}. 
{In SI Sec.~S4C and Sec.~S5D, we discuss how to obtain SJ-lines from simulations.
In particular, the finite-size analysis (see SI Fig.~S3) shows a clear difference between the SJ-lines for $\varphi_{\rm eq} =0$ (or more generally $\varphi_{\rm eq} \leq \varphi_{\rm onset}$) and $\varphi_{\rm eq} > \varphi_{\rm onset}$: only in the former case, the SJ-line becomes vertical in the thermodynamic limit~\cite{bertrand2016protocol, baity2017emergent}; 
in the latter case,   the SJ-line  is not vertical in that limit, which means that shear jamming could occur at different densities.}


{While the SJ-lines in Fig.~\ref{fig:Jplane_comparison}(a) are constructed by using strain-controlled protocols, they can be also obtained by stress-controlled protocols, which are commonly used as well to investigate the behavior of jammed systems under shearing~\cite{ciamarra2011jamming, ciamarra2009jamming}.
In stress-controlled AQS  of SSs (see SI Sec.~S7), the onset of shear jamming is signaled  by a steep increase of mechanical stress $\Sigma_{\rm mech}$ at $\gamma \approx \gamma_{\rm j}$ under the constant volume condition (see SI Fig.~S15). A similar approach has been recently used in Ref.~\cite{babu2020friction} to study shear jamming in frictionless spheres.}\\

{\bf Reversible-jamming and irreversible-jamming}

Depending on the reversibility of the route to jamming, the {whole} J-plane is divided into two regions {(see Fig.~\ref{fig:Jplane_sche} and~\ref{fig:Jplane_comparison}(b))}.

{(i) Reversible-jamming regime.} This regime contains jammed states that can be generated  by {\it reversible} routes, in the sense that the initial unjammed state at {$(\varphi_{\rm eq},0)$}
  and the final jammed state at {$(\varphi_{\rm j},\gamma_{\rm j})$}
  are in the same meta-stable glass basin {(or meta-basin)}
  and therefore the initial memory is kept {(see the illustration in Fig.~\ref{fig:Jplane_sche})}.
 {In this regime, the quench to jamming}  corresponds to the so-called {\it state-following} dynamics in structural~\cite{RUYZ14,rainone2016following} and spin~\cite{franz1995recipes,barrat1997temperature,krzakala2010following,krzakala2013performance,folena2020rethinking} glasses. {The terminology ``state-following'' is used here to emphasize the deep connection between the final jammed state and the initial equilibrium state, as the former is followed from the latter within the same meta-basin.}
  {Although the Gardner transition~\cite{charbonneau2014fractal},
    which takes place before reaching jamming,
      induces splitting of the meta-basin into many marginally stable
      sub-basins, it does not make
      the state-following dynamics irreversible as demonstrated in
      \cite{jin2018stability}.
    }

{(ii) Irreversible-jamming regime}. In this regime the routes to jamming are irreversible, in the sense that the initial unjammed state and the final shear jammed state are {in different meta-basins, or such a meta-basin can not be defined for the initial state (see the illustration in Fig.~\ref{fig:Jplane_sche})}.
  The memory of the initial condition is partially or completely lost.
{Very importantly, the J-point belongs to this regime.}


{The reversibility is determined numerically by examining the difference between the initial (unjammed) configuration before applying the jamming protocol and the final (unjammed) configuration after reverting the jamming route (see Fig.~\ref{fig:Jplane_comparison}b), quantified by the relative mean square displacement (RMSD) $\Delta_{\rm r}$ (see SI Sec.~S6 for details).}
The two regimes are separated by two  \red{crossover} lines: 
{a {\it state-following line} 
that separates state-following and non-state-following dynamics (see SI Sec.~S6C),
and a {\it yielding-jamming separation line}  formed by state points $\{\varphi_{\rm c}(\varphi_{\rm eq}), \gamma_{\rm c}(\varphi_{\rm eq})\}$, which separates shear yielding and shear jamming in thermal HSs~\cite{urbani2017shear, jin2018stability, PhysRevE.100.032140} (see SI Sec.~S6D).}
 The end point of the state-following line is the {\it state-following jamming point} at $\{\varphi_{\rm J}^{\rm SF}, 0\}$, which is compression quenched from the {\it state-following density} $\varphi_{\rm SF}$, i.e., $\varphi_{\rm J}^{\rm SF} = \varphi_{\rm IJ} (\varphi_{\rm eq} = \varphi_{\rm SF})$. {For our model, $\varphi_{\rm J}^{\rm SF}  \approx 0.67$ and  $\varphi_{\rm SF}  \approx 0.60$  (see SI Sec.~S5B and Table~\ref{table:densities}). The state-following density  $\varphi_{\rm SF} \approx 0.60$ is very close to the dynamical glass transition density $\varphi_{\rm d}  = 0.594(1)$~\cite{berthier2016growing}, both of which correspond to crossovers rather than sharp phase transitions in finite dimensions. \red{According to  the mean-field theory, $\{\varphi_{\rm c}(\varphi_{\rm eq}), \gamma_{\rm c}(\varphi_{\rm eq})\}$ is a critical point in large dimensions~\cite{urbani2017shear}, but simulation results suggest that this criticality is absent in three dimensions~\cite{jin2018stability}. Overall, the reversible and irreversible regimes are separated by gradual crossovers, rather than reversible–irreversible transitions~\cite{corte2008random, das2020unified, nagasawa2019classification}.}}

{It should be noted that, the reversibility is a property of free-energy landscape associated to the jammed state. The route to a jammed state in the reversible-jamming/irreversible-jamming regime is expected to be reversible/irreversible under any jamming protocols. 
\red{Indeed, numerical results (see Fig. S9) confirm that, in the reversible-jamming regime, the relationship between the isotropic jamming density $\varphi_{\rm IJ}$ and the initial equilibrium density $\varphi_{\rm eq}$ is independent of  the jamming protocol (athermal or thermal), as well as the compression rate used in the thermal protocol.}
However, not every protocol can explore the entire J-plane. In fact, the thermal protocols can only access mainly the reversible part as shown in Fig.~\ref{fig:Jplane_comparison} (see SI Sec.~S5 for details). This is because in order to obtain low density packings, one needs to apply rapid TC, but if the compression is too fast, the generated packings become hypostatic ($z_{\rm j}<2d$, see SI Fig.~S8).}

So far, we have discussed how to construct the J-plane and the reversibility of the routes to jamming. Below  we analyze in detail the properties of packings on the jamming-plane. \\



\begin{table}[h]
\caption{Summary of relevant densities. The table  summarizes the values of  the dynamical glass transition crossover density $\varphi_{\rm d}$, the onset density $\varphi_{\rm onset}$ of glassy dynamics, the J-point density $\varphi_{\rm J}$, the minimum isotropic jamming density $\varphi_{\rm IJ}^{\rm min, th}$ obtained by the thermal protocol, the state-following density $\varphi^{\rm SF}$, and the state-following jamming density  $\varphi_{\rm J}^{\rm SF}$.}
\begin{tabular}{ c  c c c c c  }
\hline
\hline
$\varphi_{\rm d}$~\cite{berthier2016growing}  & $\varphi_{\rm onset}$~\cite{berthier2017configurational} & $\varphi_{\rm J}$ & $\varphi_{\rm IJ}^{\rm min, th}$ &  $\varphi^{\rm SF}$ & $\varphi_{\rm J}^{\rm SF}$ \\
 \hline
 0.594(1) &  0.56 &  0.655(1) & 0.665 &  0.60 & 0.67 \\
 \hline
\hline
\end{tabular}
\label{table:densities}
\end{table}

\begin{figure}[ht]
\centerline{\includegraphics[width=1.\columnwidth]{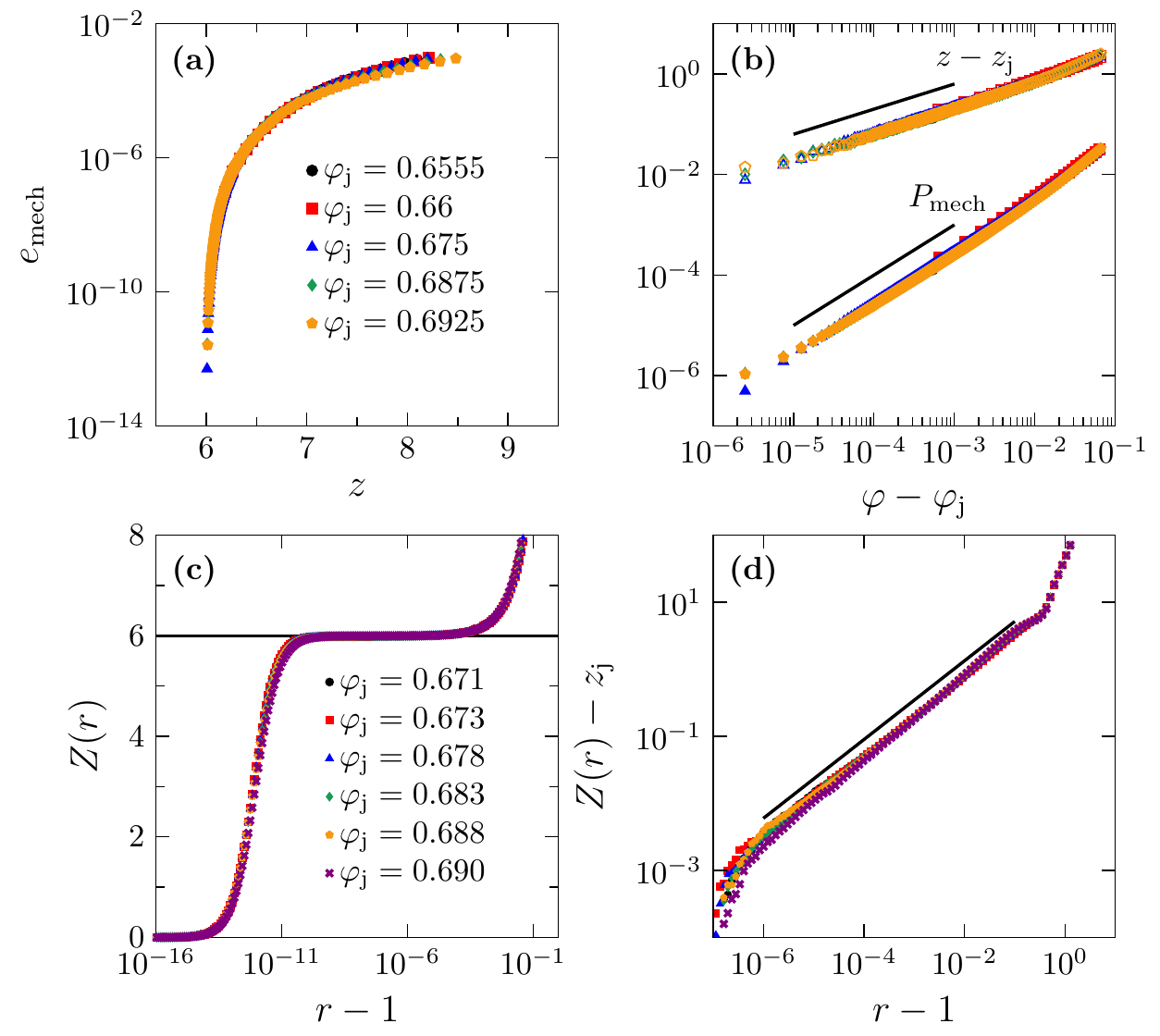}} 
\caption{Isostaticity and universality of jamming.
  We show scalings (a-b) above jamming $\varphi > \varphi_{\rm j}$ in athermal SSs,  and (c-d) below jamming $\varphi < \varphi_{\rm j}$ in thermal HSs, for $\varphi_{\rm eq} = 0.643$ and $N=8000$. 
  Data in (a-b) are obtained from athermal compressions of SS packings from  $\varphi_{\rm j}$ to $\varphi$, for a few different $\varphi_{\rm j}$ along the SJ-line
  {(blue open triangles in Fig.~\ref{fig:Jplane_comparison} and SI  Fig.~S7).}
(a) {Mechanical energy} density $e_{\rm mech}$ versus coordination number $z$.
(b) {Mechanical pressure} $P_{\rm mech}$ and the excess coordination number $z - z_{\rm j}$ as functions of $\varphi - \varphi_{\rm j}$.
  The data are consistent with scalings Eqs.~(\ref{eq:scaling_above_p}) and~(\ref{eq:scaling_above_z}) (lines). (c) Cumulative structure function $Z(r)$ of HSs below jamming, for a few different $\varphi_{\rm j}$ along the SJ-line
{(blue filled triangles in Fig.~\ref{fig:Jplane_comparison} and SI Fig.~S12)}.
The black line indicates the isostatic coordination number $z_{\rm j} = 6$. (d) The  data of $Z(r)$ is consistent with the critical jamming scaling Eq.~(\ref{eq:scaling_below_Z}) (line).} 
\label{fig:universality}
\end{figure}


{\bf  Isostaticity and jamming universality}

Any state point $\{ \varphi_{\rm j}, \gamma_{\rm j}\}$ on the J-plane corresponds to {a critical jamming state, right at a jamming-unjamming transition}~\cite{o2003jamming}. 
{Nearest-neighbor particles are just in touch in a critical jammed packing.}
Keeping the strain  $\gamma_{\rm j}$ unchanged,  the packing becomes over-jammed {when it is compressed} from $\varphi_{\rm j}$ to $ \varphi =  \varphi_{\rm j} + \delta \varphi$, where $\delta \varphi > 0$, and {becomes unjammed when it is decompressed} from   $\varphi_{\rm j}$ to $ \varphi =  \varphi_{\rm j} - \delta \varphi$.
{Quite remarkably,} as shown below, the isotropically compression jammed  and shear jammed states are all isostatic, and 
belong to  the same  universality class, i.e., 
the  jamming critical  exponents near $\varphi_{\rm j}$, from both below and above jamming,  are independent of $\gamma_{\rm j}$ (including the case $\gamma_{\rm j} = 0$).




Let us first show that the coordination number  at the {jamming-unjamming} transition satisfies the isostatic condition.
We compress the SS packings athermally from $\{\varphi_{\rm j}, \gamma_{\rm j}\}$ to $\{\varphi > \varphi_{\rm j}, \gamma_{\rm j}\}$, keeping the shear strain $\gamma_{\rm j}$ unchanged,
{and} then  measure the coordination number $z$ (without rattlers) as a function of $\varphi$. 
Figure~\ref{fig:universality}(a) shows that
the coordination number $z$ satisfies  the isostatic condition $z =  z_{\rm j} =6$, at the unjamming transition where {the mechanical energy density vanishes, $e_{\rm mech} = E_{\rm mech}/N \to 0$,} from above jamming.
Consistently, Fig.~\ref{fig:universality}(c)
shows that the isostatic condition  also holds at the jamming transition where {the reduced entropic pressure diverges, $p_{\rm entro} \to \infty$,} from below jamming in thermal HSs. Moreover, the isostatic condition is valid for any packing along the SJ-line, independent of  $\gamma_{\rm j}$ (Fig.~\ref{fig:universality}(a) and (c)). 

Second, we show that 
 all packings above jamming ($\varphi > \varphi_{\rm j}$) follow the  same set of scaling laws.
Under athermal compressions,  the {mechanical} pressure $P_{\rm mech}$, the {mechanical}  energy $E_{\rm mech}$ and the coordination number $z$ all increase in SS packings. 
Figure~\ref{fig:universality}(a) shows that, for packings along the SJ-line with the same $\varphi_{\rm eq} = 0.643$ but different $\varphi_{\rm j}$ 
{(blue open triangles in Fig.~\ref{fig:Jplane_comparison}(a)),}
the data of energy density $e_{\rm mech}$ versus $z$ collapse onto the same master curve, which vanishes at $z_{\rm j} =6$. 
Furthermore, the following  scalings,
which are well known for isotropically jammed packings~\cite{o2003jamming},
 are also satisfied  in shear jammed packings (Fig.~\ref{fig:universality}b):
\beq
P_{\rm mech} \sim \varphi - \varphi_{\rm j},
\label{eq:scaling_above_p}
\eeq 
and 
\beq
z - z_{\rm j} = (\varphi - \varphi_{\rm j})^{1/2}.
\label{eq:scaling_above_z}
\eeq

Third, we examine the scaling behavior below jamming ($\varphi < \varphi_{\rm j}$).
To do that, we compute the cumulative structure function $Z(r)$ of HS packings along the SJ-line for $\varphi_{\rm eq} = 0.643$. The packings are compressed or sheared until the pressure reaches   
 $p_{\rm entro} = 10^{12}$. The  cumulative structure function is defined as 
\beq
Z(r) = \rho  \int_0^{r} ds 4 \pi s^{2} g(s),
\eeq
where $g(s)$ is the pair correlation function,
\beq
g(s) = \frac{1}{4 \pi s^2 \rho N} \left \langle \sum_{i\neq j} \delta\left(s - {r_{ij}/D_{ij}} \right) \right \rangle,
\eeq
with $\delta(x)$ being the delta function, {$\rho$ the number density, $D_{ij} = (D_i+D_j)/2$ the average diameter, and $r_{ij}$ the inter-particle distance}.
The cumulative structure function $Z(r)$ exhibits a plateau at $z_{\rm j}=6$  (Fig.~\ref{fig:universality}c).
The growth from this plateau satisfies the scaling 
\beq
 Z(r) - z_{\rm j} \sim (r-1)^{1-\alpha},
 \label{eq:scaling_below_Z}
\eeq
where $\alpha = 0.41269$~\cite{charbonneau2014fractal}. This scaling is predicted by the mean-field theory and has been verified numerically for isotropically jammed packings in finite dimensions~\cite{charbonneau2014fractal}. Here we show that it also holds for shear jammed packings. \\


{\bf  Anisotropy of contact networks}

We use the  anisotropy parameter $R_{\rm A}$, which is based on the fabric tensor $\hat{R}$,  to quantify the anisotropy of contact networks in jammed packings. 
 The fabric tensor is defined as 
\beq
\hat{R} = \frac{1}{N} \sum_{i \neq j}  \frac{\vec{r}_{ij}}{|\vec{r}_{ij}|}  \otimes  \frac{\vec{r}_{ij}}{|\vec{r}_{ij}|}, 
\eeq
where $\vec{r}_{ij}$ is the vector connecting two particles $i$ and $j$ that are in contact, and $\otimes$ denotes a vector outer product. The eigenvalues of  $\hat{R}$  are denoted by $\lambda_1$, $\lambda_2$ and  $\lambda_3$, and the coordination number is related to the eigenvalues by $z_{\rm j}= \lambda_1+\lambda_2+\lambda_3$. The fabric anisotropy parameter $R_{\rm A}$ is defined as the difference between the largest and the smallest eigenvalues, normalized by $z_{\rm j}$, $R_{\rm A} = (\lambda_{\rm max} - \lambda_{\rm min})/z_{\rm j}$~\cite{vinutha2016disentangling}.

\begin{figure}[ht]
\centerline{\includegraphics[width=1.0\columnwidth]{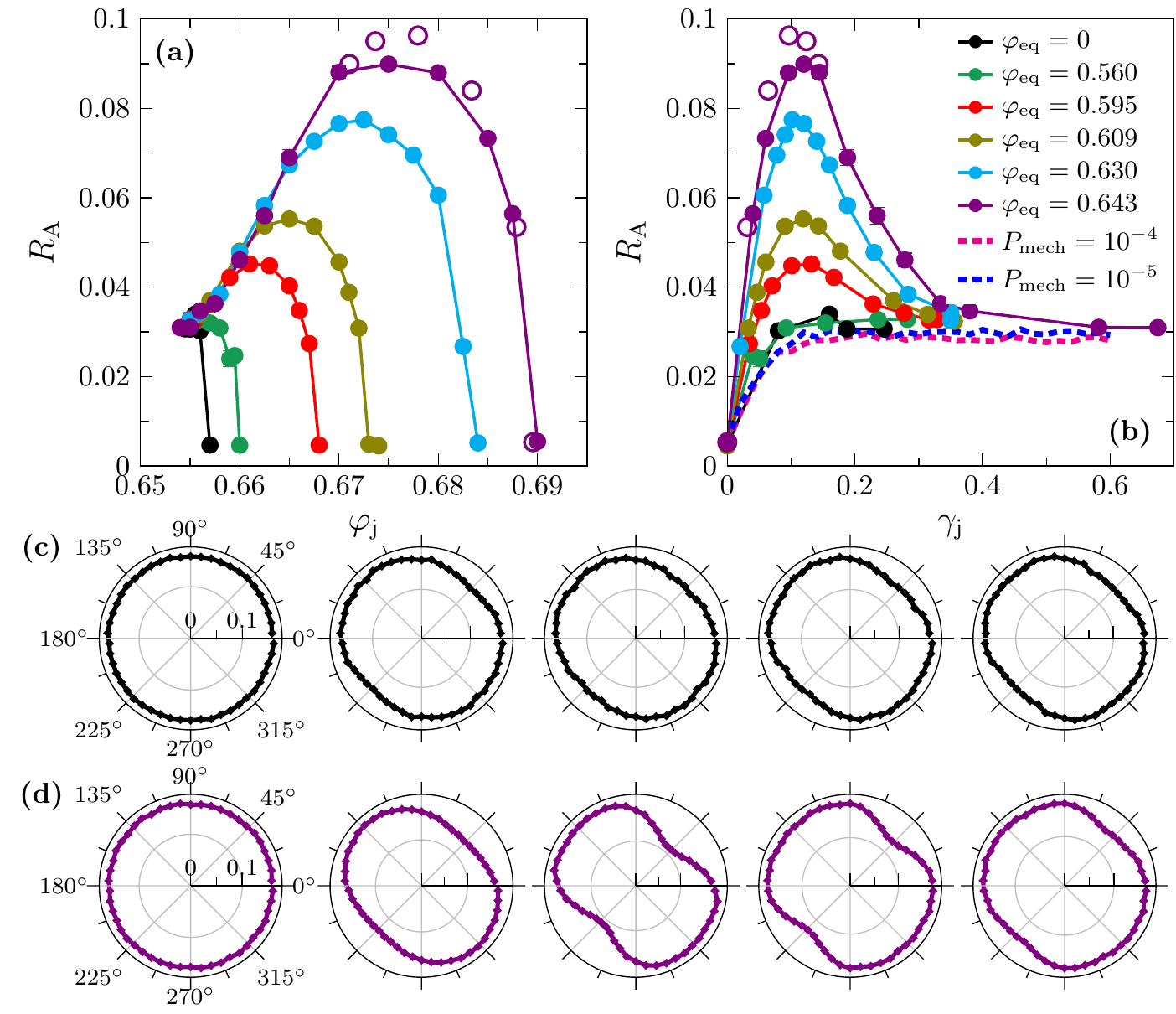}} 
\caption{Contact anisotropy. Fabric anisotropy parameter $R_{\rm A}$ of  packings along SJ-lines, as a function of (a) jamming density $\varphi_{\rm j}$ and (b) jamming strain $\gamma_{\rm j}$. Data are obtained for $N=8000$ and a few different $\varphi_{\rm eq}$,  by using the athermal protocol (filled symbols) and the thermal protocol (open symbols). 
\redd{Dashed lines in (b) represent data obtained for the critical state, i.e. steady states in AQS simulations under constant, vanishingly small pressures ($P_{\rm mech} = 10^{-4}$ and $10^{-5}$, $N=8000$, averaged over 48 samples) }
(c) Contact angle probability distribution $P_\theta(\theta)$ for $\varphi_{\rm eq}=0$ and $\gamma_{\rm j} = 0, 0.08, 0.16, 0.19, 0.25$ (from left to right). (d) Contact angle probability distribution $P_\theta(\theta)$ for $\varphi_{\rm eq}=0.643$ and   $\gamma_{\rm j} = 0, 0.04, 0.12, 0.28, 0.37$ (from left to right).
 }
\label{fig:anisotropy}
\end{figure}

Another important quantity to characterize the  anisotropy is the contact angle probability distribution $P_\theta(\theta)$, where the angle $\theta$ is defined through the coordinate transformation $\vec{r}_{ij} = (r_{ij} \sin \theta \sin\phi, r_{ij}\cos\phi, r_{ij} \cos\theta \sin \phi)$, considering that the shear strain is applied in the $x-z$ plane. 
From the lowest order Fourier expansion,   $P_\theta(\theta)$ is   related to the fabric anisotropy parameter $R_{\rm A}$ via~\cite{radjai1998bimodal},
\beq
P_\theta(\theta) \approx \frac{1}{2 \pi} \left[ 1+2R_{\rm A} \cos 2\left( \theta - \theta_{\rm c}\right)  \right],
\eeq
where $\theta_{\rm c}$ is the principle direction. 

Apparently, for isotropic jamming ($\gamma_{\rm j} = 0$), the  fabric anisotropy parameter $R_{\rm A}$ should be nearly zero, and the contact angle distribution should be uniform, as confirmed in Figs.~\ref{fig:anisotropy}(a) and (c). For shear jamming, let us consider two cases. 
In the case {of}  $\varphi_{\rm eq}=0$, the  fabric anisotropy parameter immediately jumps to a finite value $R_{\rm A} \approx 0.03$ for  non-zero $\gamma_{\rm j}$, and stays as a constant for larger $\gamma_{\rm j}$ (Fig.~\ref{fig:anisotropy}(a) and (b)),
which is consistent with the observation in \cite{chen2018stress} (Ref.~\cite{chen2018stress} also suggests that $R_{\rm A}(\gamma_{\rm j})$ would jump discontinuously at $\gamma_{\rm j}=0$
 in the thermodynamic limit).
Accordingly,  the contact angle distribution $P_\theta(\theta)$,  with  $\theta_{\rm c} = 135^{\circ}$, also quickly converges to the asymptotic distribution (Fig.~\ref{fig:anisotropy}(c)). 
In the case {of} $\varphi_{\rm eq} = 0.643$, which is above $\varphi_{\rm onset}$, the fabric anisotropy does not change monotonically with $\gamma_{\rm j}$ (Fig.~\ref{fig:anisotropy}(a) and (b)). At intermediate  $\gamma_{\rm j}$, the $P_\theta(\theta)$ has a dumbbell shape, indicating a strong  anisotropy (Fig.~\ref{fig:anisotropy}(d)). At larger $\gamma_{\rm j}$, both $R_{\rm A}$ and  $P_\theta(\theta)$ converge to the same asymptotic behaviors as in the case of $\varphi_{\rm eq}=0$. \\

\redd{It is interesting to compare the anisotropy of jammed states on the J-plane to that  of critical state. Following Ref.~\cite{babu2020friction}, steady states are obtained by constant pressure AQS of SSs at large shear strains ($\gamma \gg \gamma_{\rm Y}^{\rm mech} \sim 0.1$) where  sample-averaged physical quantities do not change anymore with the strain. The constant pressure simulation is realized by minimizing the enthalpy after each strain step~\cite{babu2020friction}.
By definition, the critical state is a steady state in the zero pressure limit $P_{\rm mech} \to 0$. 
In Fig.~\ref{fig:anisotropy}(b), we present $R_{\rm A}$ for two small pressures $P_{\rm mech} = 10^{-4}$ and  $10^{-5}$, where initial configurations are rapidly quenched from $\varphi_{\rm eq} = 0$.
At large $\gamma$, $R_{\rm A}$ (as well as other quantities such as the bond-orientational order parameter plotted in Fig.~\ref{fig:properties}(b)) approaches  a constant with a negligible pressure dependence, which means that the system has reached the critical state within the numerical precision. 
Special attention is paid to (i) the generic compression jammed state at the J-point $\{\varphi_{\rm j}  = \varphi_{\rm J} = 0.655(1), \gamma_{\rm j} = 0 \}$, (ii) the asymptotic shear jammed state at $\{\varphi_{\rm j}  = \varphi_{\rm J}, \gamma_{\rm j} \to \infty \}$, and (iii) the critical state at $\varphi_{\rm c} = 0.656(1)$. We find that $\varphi_{\rm J} = \varphi_{\rm c} $ within numerical precision, as has been already shown in~\cite{babu2020friction}, and the asymptotic shear jammed state and the critical state have a common, non-zero degree of anisotropy $R_{\rm A} \approx 0.03$, while the J-point state is clearly isotropic with  $R_{\rm A} \approx 0$.} 

{\bf Bond-orientational order}

Although the crystalline order is absent in our polydisperse model,
the packings could have structures at the local scale~\cite{ozawa2017exploring}. We characterize such order 
by using the weighted bond-orientational order parameter~\cite{mickel2013shortcomings}, defined as 
\beq
Q_{l, i}^{\rm w} = \sqrt{\frac{4 \pi}{2l + 1} \sum_{m=-l}^l \left| \sum_{j=1}^{n_{\rm b}(i)} \frac{A_{ij}}{A_i} Y_{l,m}(\vr_{ij})\right|^2},
\eeq
where $Y_{l,m}(\vr_{ij})$ is the spherical harmonic of degree $l$ and order $m$, $A_{ij}$ is the area of the Voronoi cell face between particles $i$ and $j$, and $A_i = \sum_j A_{ij}$. Here we consider the average bond-orientational order parameter with $l = 6$:
\beq
Q_{6}^{\rm w} = \frac{1}{N} \sum_i Q_{6, i}^{\rm w}.
\eeq
Figure~\ref{fig:properties}(a) shows the $Q_{6}^{\rm w}-\varphi_{\rm j}$ order map~\cite{torquato2000random} obtained from the packings associated to the J-plane. The order parameter $Q_{6}^{\rm w}$ of isotropically jammed packings ($\gamma_{\rm j} = 0$) increases with $\varphi_{\rm eq}$. Ref.~\cite{coslovich2017exploring} has shown  that this increase is inherited from the initial equilibrium  configurations at $\varphi_{\rm eq}$.
 Along SJ-lines, the $Q_{6}^{\rm w}$ decreases with decreasing $\varphi_{\rm j}$, or increasing $\gamma_{\rm j}$ (Fig.~\ref{fig:properties}b).
Interestingly, our result shows that the J-point has the minimum order $Q_{6}^{\rm w}$, and therefore the maximum randomness, among all state points on the J-plane. In this sense, the J-point coincides with the MRJ point introduced in Ref.~\cite{torquato2000random}.
 However, we emphasize that the crystalline order is excluded from our consideration, which is an essential difference from  Ref.~\cite{torquato2000random}.
 \redd{Furthermore, we find that the isotropic MRJ state (J-point state), the asymptotic shear jammed state, and the critical state all display the same  $Q_{6}^{\rm w}$ (Fig.~\ref{fig:properties}).} \\

\begin{figure}[ht]
\centerline{\includegraphics[width=1.0\columnwidth]{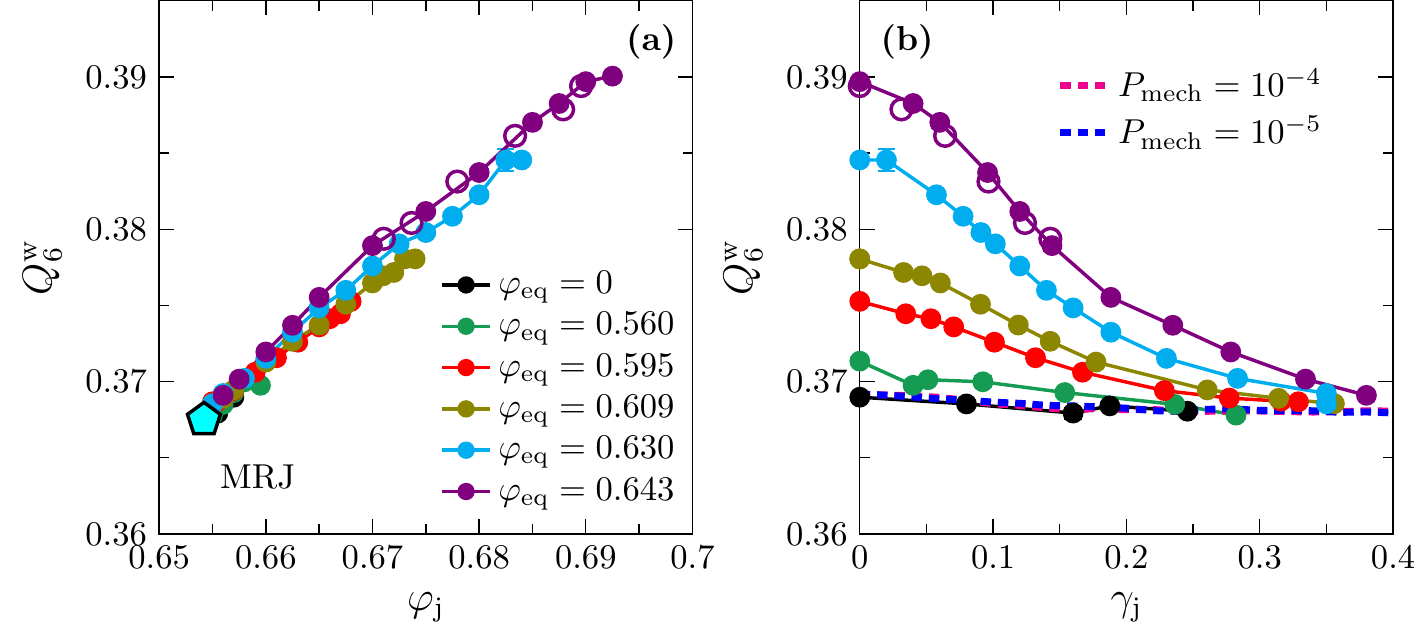}} 
\caption{{Order map.} The weighted bond-orientational order parameter $Q_6^{\rm w}$ is plotted as a function of (a) $\varphi_{\rm j}$ and (b) $\gamma_{\rm j}$, along SJ-lines with different $\varphi_{\rm eq}$\redd{, for N=8000 systems.}  
\redd{Dashed lines in (b) represent data obtained for the steady state in AQS simulations
under constant, vanishingly small pressures (lines for $P_{\rm mech} = 10^{-4}$ and $P_{\rm mech} = 10^{-5}$ fall on top of each other).}
}
\label{fig:properties}
\end{figure}

{\bf \Large Discussion}

In this paper, the concept of J-plane
is introduced,
 and is realized numerically using athermal and thermal protocols. 
Thanks to the  swap algorithm, we are able  to explore the J-plane over a wide range of jamming densities. 
It is possible to replace the role of the swap algorithm  by other protocols that  are easier to be reproduced in experiments, {such as the mechanical training protocols used in~\cite{kumar2016memory, das2020unified} }
Indeed, similar SJ-lines as in SI Fig.~S7 have been obtained in Ref.~\cite{kumar2016memory} ({see Fig.~6b there}), although within a much narrower range of densities. 
 We therefore expect the J-plane to be reproducible in tapping and shear experiments of granular matter.

Our analysis reveals  that the J-point is a rather special point on the J-plane. The state at the J-point has the minimum packing density and the maximum randomness among all possible amorphous, frictionless, jammed states. 
The remaining challenge {to theories} is to provide a first-principle understanding of the J-point. Our results  show that  $\varphi_{\rm J} \neq \varphi_{\rm IJ}(\varphi_{\rm eq} = \varphi_{\rm d})$, which {rules} out the possibility that the state at the J-point is followed from the equilibrium state at the dynamical glass transition density $\varphi_{\rm d}$.
Instead, the  recent spin-glass theory of quench dynamics~\cite{castellani2005spin}, together with earlier numerical studies~\cite{ozawa2012jamming},  suggest that
 $\varphi_{\rm J} = \varphi_{\rm IJ}(\varphi_{\rm eq} \leq \varphi_{\rm onset})$. 
Therefore {the} generalization of  the calculation in  Ref.~\cite{castellani2005spin} to sphere systems would be very appealing.

We show that the phase space of jammed states can be significantly extended by adding shear, which introduces anisotropy to the contact networks. 
Our results 
{disprove}
the earlier understanding  that shear jamming and isotropic jamming always occur at the  same jamming density in the thermodynamic limit~\cite{bertrand2016protocol, baity2017emergent}. The reversibility of the routes to jamming has a deep connection to the reversibility of the corresponding thermal HS glasses upon quench or shear~\cite{jin2018stability}. 

{The J-plane proposed here shall not be  confused  with the zero-temperature Liu-Nagel jamming phase diagram~\cite{liu1998nonlinear}.
The latter is defined by a yield stress line, which separates jammed and unjammed regions on the stress-density plane (note that this stress  refers to the mechanical stress of athermal SS systems, which is different from the entropic stress of HSs, see SI Sec.~S1), while the former is the collection of all jammed states on the jamming strain-jamming density plane, for which the  mechanical stress is always zero. In principle, each state point $\{\varphi_{\rm j}, \gamma_{\rm j}\}$ on the J-plane can be extended into a  Liu-Nagel-like phase diagram, as follows.
A system at $\{\varphi_{\rm j}, \gamma_{\rm j}\}$ can be compressed into  an over-jammed state at $\{\varphi > \varphi_{\rm j}, \gamma_{\rm j}\}$, whose mechanical yield stress  is a function of $\varphi$.
This yield stress-density line gives a generalized  Liu-Nagel phase diagram, with the original version~\cite{liu1998nonlinear} corresponding to the special case  $\{\varphi_{\rm j} = \varphi_{\rm J}, \gamma_{\rm j}=0\}$.
Indeed, very recently such a generalization has been done for the cases $\{\varphi_{\rm j} > \varphi_{\rm J}, \gamma_{\rm j} = 0\}$, 
 where the onset of yield stress becomes discontinuous at jamming~\cite{babu2020friction}.
}

{Interestingly, frictionless shear jamming can be also observed in crystals such as FCC lattices (see SI Sec.~S8). 
Because  crystals are in equilibrium, accordingly the SJ-line becomes unique and independent of $\varphi_{\rm eq}$. The separation between reversible-jamming and irreversible-jamming, and its connection to yielding of thermal HSs, remain to be present (see SI Fig.~S16).}

Finally, 
\redd{our results should pave the way}
for a set of novel studies. 
Related open questions include, but are not limited to:  
do the jammed states on the J-plane share the same rheological properties, before and after yielding~\cite{jin2017exploring, jin2018stability}? 
How to extend the J-plane in order to integrate the effects of friction~\cite{bi2011jamming, vinutha2016disentangling, otsuki2011critical} and crystalline order~\cite{kapfer2012jammed}? 
Can we make a  connection between reversible-jamming/irreversible-jamming discussed here and 
the reversible-irreversible transition in suspensions~\cite{corte2008random} and granular systems~\cite{das2020unified, nagasawa2019classification}?\\

\centerline{\bf METHODS}

Full materials and methods are included in SI, where we describe in detail the simulation models  (Sec.~S1), the swap algorithm that is used to prepare initial configurations (Sec.~S2), and the protocols to obtain jammed configurations (Secs.~S3 and~S7). We also explain how to explore the J-plane using athermal (Sec.~S4) and thermal jamming protocols (Sec.~S5), and how to analyze the reversibility of routes to jamming (Sec.~S6). More general jamming protocols are discussed in Sec.~S9.


\begin{acknowledgments}
We warmly thank A.~Altieri, M.~Baity-Jesi, B.~Chakraborty, P. Charbonneau, S.~Chen, G.~Folena, H.~Hayakawa, T.~Kawasaki,  K.~Miyazaki, C.~O'Hern, M.~Otsuki, D.~Pan, S.~Sastry, S.~Teitel, P.~Urbani, Y.~Wang, J.~Zhang, and F.~Zamponi for discussions.
  This work was supported by KAKENHI (No. 25103005  ``Fluctuation \& Structure'', No.~19H01812 and No.~20H00128) from MEXT, Japan. Y.J. acknowledges funding from Project 11974361,  Project 11935002, and Project 11947302 supported by NSFC, from Key Research Program of Frontier  Sciences, CAS, Grant NO. ZDBS-LY-7017, and from the CAS Pioneer Hundred Talents Program. 
  The computations were performed using the computing facilities in Research Center for Computational Science, Okazaki, Japan, the  computing facilities  in 
  the Cybermedia center, Osaka University, the HPC Cluster of ITP-CAS, and Tianhe-2 Supercomputer, National Supercomputer Center in Guangzhou.
\end{acknowledgments}

 \clearpage

\onecolumngrid

\centerline{\bf Supplementary Information}

\setcounter{figure}{0}  
\setcounter{equation}{0}  
\setcounter{table}{0} 
\renewcommand\thefigure{S\arabic{figure}}
\renewcommand\theequation{S\arabic{equation}}
\renewcommand\thesection{S\arabic{section}}
\renewcommand\thetable{S\arabic{table}}

\section{Models} 
\label{sec:model}
In experiments, 
colloidal suspensions and granular matter are two typically studied  jamming systems.
A colloidal suspension jams when the viscosity diverges~\cite{boyer2011unifying, peters2016direct},
while the jamming of granular matter occurs at the onset of rigidity~\cite{bi2011jamming, aste2005geometrical, coulais2014shear}. 
The main difference between the two systems lies in the motion of particles. 
Granular particles are large enough such that their thermal motions can be neglected,
{and therefore the systems are athermal by nature.}
To numerically model  suspensions of (hard) colloidal particles, 
we simulate thermal hard spheres (HSs) following Newtonian molecular dynamics ~\cite{torquato2000random, charbonneau2011glass}. To model granular materials, we simulate 
harmonic soft spheres (SSs) following overdamped quasistatic dynamics~\cite{o2003jamming}. We neglect the effects of  friction~\cite{bi2011jamming, vinutha2016disentangling, seto2019shear}, adhesion~\cite{liu2017equation}, and hydrodynamic interactions~\cite{seto2019shear}, and use models with a large polydispersity to prevent crystallization~\cite{berthier2017configurational, ozawa2017exploring, BCNO2016PRL}.

{Our} systems consist of $N$ spherical particles in a simulation box of volume $V$, for a continuous diameter distribution $P(D) \sim D^{-3}$, where $D_{\rm min} \leq D \leq D_{\rm min}/0.45$~\cite{BCNO2016PRL}.  The number density is   $\rho=N/V$  and the volume fraction is
$\varphi=\rho (4/3)\pi \overline{D^3}$. The mean diameter $\overline{D}$ is set as the unity of length, and all particles have the same unit mass $m=1$. For the same $P(D)$, two models are studied.

\begin{itemize}
\item[(i)] {\bf Thermal hard sphere model}. 
The model represents  a suspension of hard colloidal particles  with negligible friction.
The simulation is performed under  constant unit temperature $T=1$. Because the potential energy is always zero and only inter-particle collisions contribute, both pressure $P_{\rm entro}$ and stress $\Sigma_{\rm entro}$ are purely {\it entropic}. We define the reduced entropic pressure as $p_{\rm entro}=P_{\rm entro}V/Nk_{\rm B}T$ and the reduced entropic stress as $\sigma_{\rm entro}=\Sigma_{\rm entro} V/Nk_{\rm B}T$, and set the  Boltzmann
constant $k_{\rm B} = 1$. The method to compute $p_{\rm entro}$ and $\sigma_{\rm entro}$ is explained in detail in Ref.~\cite{jin2017exploring}.
Jamming occurs when the entropic pressure and stress diverge.  

\item[(ii)]{\bf Athermal soft sphere model}. The model represents a frictionless granular system. 
The SS potential has a harmonic form $U(r_{ij}) = \frac{1}{2} (1- r_{ij}/D_{ij})^2$ (zero if $r_{ij}>D_{ij}$), where $r_{ij}$ is the inter-particle distance between particles $i$ and $j$, and $D_{ij} = (D_i + D_j)/2$ is the mean diameter. The simulation is performed at zero temperature. Both pressure $P_{\rm mech}$ and stress $\Sigma_{\rm mech}$ are contributed by the mechanical  contacts between particles, and therefore are purely  
 {\it mechanical}. 
 The SSs are jammed if  $P_{\rm mech} > 0$, and the unjamming transition occurs as $P_{\rm mech} \to 0$.

 \end{itemize}
 

\section{{Preparation of  initial configurations using a swap algorithm}}
\label{sec:initial}
To be used as the 
initial states, we generate 
equilibrium {liquid} configurations of the  thermal HS model  at $\varphi_{\rm eq} > 0$, as well as  
purely random (ideal gas) configurations at $\varphi_{\rm eq} = 0$. 
Note that, the HS liquid configuration at the low-density limit is identical to the ideal gas configuration, because the excluded volume effect is negligible. 
{To use these initial configurations in the athermal protocol ({Sec.} \ref{sec:athermal}), we simply need to replace the thermalized HSs by SSs at the same positions, and switch off the temperature.}

The initial configurations {with $\varphi_{\rm eq}>0$}
  are prepared using a swap algorithm~\cite{BCNO2016PRL}. 
At each swap Monte Carlo  step, two randomly chosen particles are swapped if they do not overlap with other particles at the new positions.
The swap moves, {combined} with event-driven molecular dynamics~\cite{jin2017exploring, jin2018stability}, 
significantly {facilitate} the equilibration procedure. The diameter distribution $P(D)$ (see {Sec.}~\ref{sec:model}) was designed to optimize the efficiency  of the swap algorithm~\cite{ninarello2017models}. This algorithm allows us to equilibrate HS systems over a wide range of volume densities, $\varphi_{\rm eq} \in [0, 0.655]$. For several  chosen $\varphi_{\rm eq}$ in this range, we prepare equilibrium configurations {with} a few different system sizes $N=250, 500, 1000, 2000, 4000, 8000$. For each $N$, 500-2000 independent samples are generated, in order to obtain sufficient statistics. 


The equilibrium  liquid equation of state (EOS) $p_{\rm entro}^{\rm L}(\varphi_{\rm eq})$ can be well described  by the empirical  Carnahan-Stirling form~\cite{berthier2016growing} (see Fig.~\ref{fig:athermal_protocol}). The non-equilibrium glass EOS $p_{\rm entro}^{\rm G}(\varphi; \varphi_{\rm eq})$ depends on the glass transition density $\varphi_{\rm g} = \varphi_{\rm eq}$, where the system falls out of equilibrium~\cite{berthier2016growing}. {In practice this is obtained
by the TC protocol explained in {Sec.~\ref{sec:protocols}}:
the standard event-driven molecular dynamics without swap 
are used to simulate the compression process}.


\section{{Jamming protocols}}
\label{sec:protocols}
{We describe four jamming protocols:
(i) athermal rapid compression (ARC)~\cite{o2003jamming}, which is equivalent to rapid quench,   
(ii) {strain-controlled} athermal quasistatic shear (AQS)~\cite{bertrand2016protocol, baity2017emergent}, 
{(iii)} {thermal compression (TC)}~\cite{donev2005pair, torquato2000random}, and {(iv) strain-controlled thermal quasi-static shear (TQS)~\cite{jin2017exploring,jin2018stability}}, all of which have been commonly used in the literature. The athermal protocols (i and ii) are applied to SSs (see Fig.~\ref{fig:athermal_protocol} for an illustration), and the thermal protocols (iii and iv) are applied to HSs (see Fig.~\ref{fig:thermal_protocol} for an illustration). Both shear protocols here are strain-controlled, which are used to obtain our main results. In Sec.~\ref{sec:stress_controlled} we introduce an additional, stress-controlled AQS protocol, which gives results consistent with those obtained by strain-controlled shear protocols. }

\begin{figure}[h!]
\centerline{\includegraphics[width=0.4\columnwidth]{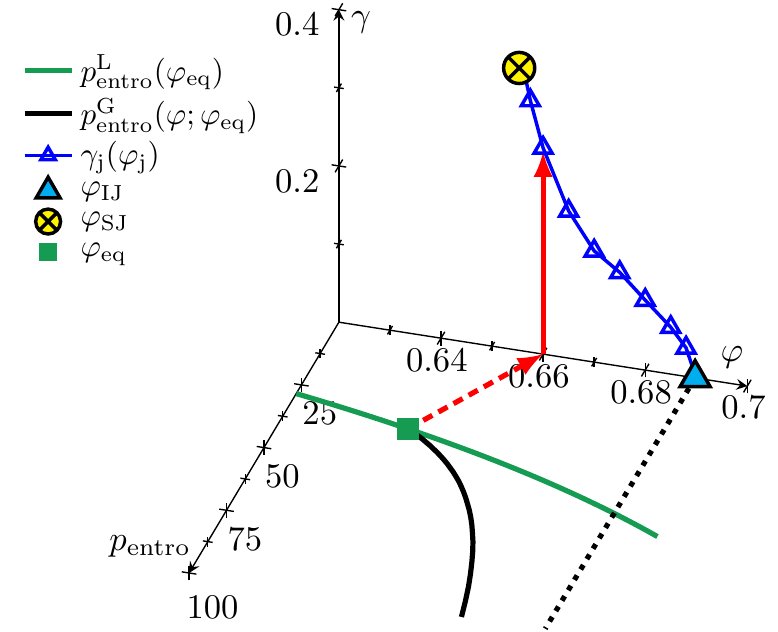}} 
\caption{{Illustration of the athermal {protocols}.} 
  The example   shows how  a shear jammed packing of $N=8000$ SSs is obtained, {starting from a thermalized initial state at density $\varphi = \varphi_{\rm eq}$ and zero strain. The {entropic} pressure and density of the initial state are related 
    by the liquid EOS
    $p_{\rm entro}^{\rm L}(\varphi_{\rm eq})$.}
 {After switching off the temperature,} the system is {athermally compressed}
  from $\varphi_{\rm eq} = 0.643$ to the target density
${\varphi}= 0.660$ {(red dotted arrow line)},
and then 
is shear jammed at
${\gamma} = 0.259$ 
by using AQS {(red solid arrow line)}. This jammed state is described by a state point $\{\varphi_{\rm j} = 0.660, \gamma_{\rm j} = 0.259 \}$ on the J-plane.
\red{Other points along the shear jamming line (SJ-line) $\gamma_{\rm j}(\varphi_{\rm j})$ are obtained
similarly starting from the same initial state (green square), but using different target densities $\varphi$.}
}  
\label{fig:athermal_protocol}
\end{figure}

\subsection{{Athermal protocols}}
\label{sec:athermal_protocol}

\begin{itemize}
\item[(i)] {\bf Athermal rapid compression}.  
First, we instantaneously switch off the temperature of a HS equilibrium configuration at $\varphi_{\rm eq}$.
We then work at zero temperature using the SS inter-particle potential.
The particle sizes are 
inflated or deflated proportionally and instantaneously to match a target density $\varphi_{\rm j}$. 
Overlaps between particles are removed by minimizing the total potential energy using the FIRE algorithm~\cite{bitzek2006structural}. The $\varphi_{\rm eq}=0$ limit of this procedure is exactly the well-known jamming algorithm introduced by O'Hern et al.~\cite{o2003jamming} to study the J-point. 

\item[(ii)] {\bf Athermal quasistatic shear}. If the configuration is not jammed after the previous step, we further apply AQS to the unjammed configuration.
A simple shear deformation in the $x$-$z$ direction  is applied under Lees-Edwards boundary conditions~\cite{lees1972computer}.  At each step, all particles are shifted instantaneously by $x_i \to x_i + \delta \gamma z_i$, where $x_i$ and $z_i$ are the $x-$ and $z-$coordinates of particle $i$, and  $\delta \gamma $ is the strain step size, 
 followed by the FIRE algorithm  to remove the overlaps.
 The AQS is stopped either when the system is jammed at  a certain jamming strain $\gamma_{\rm j}$, or the strain $\gamma$ exceeds a maximum value  $\gamma_{\rm max}$. 
 The parameter values,   $\delta \gamma  = 0.02$ and $\gamma_{\rm max} = 0.4$ (unless otherwise specified), are the same  as in Ref.~\cite{baity2017emergent}. The $\varphi_{\rm eq}=0$ limit of this procedure is equivalent to the one employed in~\cite{baity2017emergent}.
 \end{itemize}
 
{\bf Criterion of athermal jamming.} We use the same jamming criterion as in Ref.~\cite{o2003jamming}. A system is unjammed if the 
{the mechanical energy density $e_{\rm mech}$}
decays below  $10^{-16}$ after the energy minimization; it is jammed if $e_{\rm mech} > 10^{-16}$ and 
the difference $\delta e_{\rm mech}$ between successive steps is less than $10^{-15}$. The same criterion is applied to both isotropic and shear jamming.

\begin{figure}[h!]
\centerline{\includegraphics[width=0.4\columnwidth]{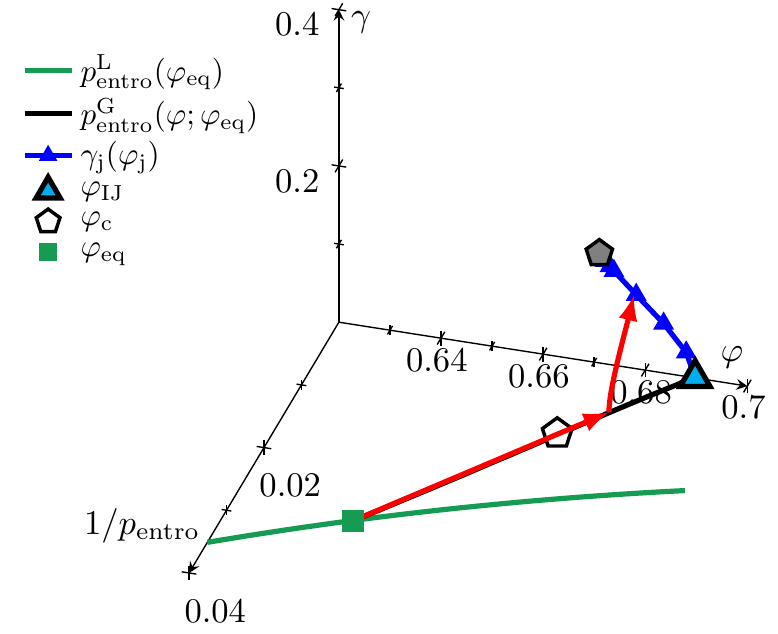}} 
\caption{Illustration of the thermal {protocols} (to be compared with the athermal {protocols}, see Fig.~\ref{fig:athermal_protocol}).
The example (red arrow line) shows how a  shear jammed packing of $N=8000$ HSs is obtained.
 The system is compressed from $\varphi_{\rm eq} = 0.643$ to the target density $\varphi_{\rm j} = 0.678$ with a constant compression rate $\Gamma = 10^{-4}$, and then is sheared  
 with a constant shear rate $\dot{\gamma} = 10^{-4}$, which results in shear jamming at strain  $\gamma_{\rm j} = 0.095$.
 }  
\label{fig:thermal_protocol}
\end{figure}

\subsection{{Thermal protocols}}
\label{sec:thermal_protocol}

\begin{itemize}
 \item[{(iii)}]{\bf Thermal compression.}
To simulate the TC procedure, we
 use the Lubachevsky-Stillinger  algorithm~\cite{skoge2006packing}, which is based on event-driven molecular dynamics.
Starting from an equilibrium configuration at $\varphi_{\rm eq}$, the  algorithm compresses HSs by inflating their sizes  with a fixed rate  $\Gamma = \frac{1}{2D} \frac{dD}{dt}$. 
The simulation time is expressed in unit of $\sqrt{1/k_{\rm B} m\overline{D}^2}$.  Although here $\Gamma$ and  $\varphi_{\rm eq}$  are treated as independent control parameters, in principle they play equivalent roles. Both of them control the glass transition density $\varphi_{\rm g}$, which eventually determines the jamming density $\varphi_{\rm IJ}$~\cite{berthier2016growing}.
If  $\varphi_{\rm eq} = 0$ and swaps are switched off (as in the conventional Lubachevsky-Stillinger  algorithm), then $\varphi_{\rm g}$ only depends on the compression rate $\Gamma$. In this case, the maximum $\varphi_{\rm g}$ is around the dynamical glass transition crossover density $\varphi_{\rm d} = 0.594$, obtained by the slowest compression  ($\Gamma \sim 10^{-6}$) that can be achieved  in our simulations. 
If the swaps are switched on, then the maximum $\varphi_{\rm g}$ depends on the maximum equilibrium density $\varphi_{\rm eq}^{\rm max} \approx 0.655$ that can be achieved by the swap algorithm, which is significantly higher than $\varphi_{\rm d}$.
For $\varphi_{\rm eq} > \varphi_{\rm d}$, the glass transition density $\varphi_{\rm g}$, as well as the jamming density $\varphi_{\rm IJ}$, mainly depends on $\varphi_{\rm eq}$ rather than on $\Gamma$ (see {Sec.~\ref{sec:thermal}} for a detailed analysis).

\item[{(iv)}]{\bf Thermal quasi-static shear.}
To simulate shear, we apply  constant volume simple shear in the $x$-$z$ direction with a fixed rate $\dot{\gamma}  = 10^{-4}$. 
For $\varphi_{\rm eq} > \varphi_{\rm d}$, the  stress-strain curves are insensitive to the variation of $\dot{\gamma}$ for a few orders of magnitude~\cite{jin2017exploring}.
At each step, we perform 1000 collisions per particle, and then instantaneously increase the shear strain by $\delta \gamma = \dot{\gamma} \delta t$, where  $\delta t$ is the time elapsed during the collisions. All particles are shifted by $x_i \to x_i + \delta \gamma z_i$.
To remove the possible 
overlappings introduced during this shift, we switch to the SS potential and use the FIRE algorithm to minimize the energy. The SS potential is switched off after.  As in the thermal protocol,  Lees-Edwards boundary conditions are used.

\end{itemize}

{\bf Criterion of thermal jamming.} 
A HS configuration is jammed if its reduced {entropic} pressure $p_{\rm entro} > 10^5$ and the average coordination number 
$z_{\rm j}$ satisfies the isostatic condition, $z_{\rm j}=2d =6$.  In the calculation of coordination number, we remove the rattlers who have less than four contacts. 
We stop shear simulations if the system reaches  the maximum strain $\gamma_{\max} = 0.2$ without jamming. This maximum strain $\gamma_{\max} = 0.2$ is greater than the typical yielding strain {$\gamma_{\rm Y}^{\rm entro} \sim 0.1$} of our model~\cite{jin2018stability}. 

\section{Exploring the jamming-plane using {athermal protocols}}
\label{sec:athermal}
 {We generate a large number of 
 isotropically jammed ({using} ARC) or shear jammed ({using} ARC + AQS)
packings,
starting from independent samples of initial configurations
at $\{\varphi_{\rm eq},0\}$.
For each of them, we precisely locate the state point $\{\varphi_{\rm j},\gamma_{\rm j}\}$
at which the system becomes jammed. 
For a given $\{\varphi_{\rm eq},0\}$, we find fluctuations of
$\{\varphi_{\rm j},\gamma_{\rm j}\}$ among  samples, which strongly
{depend} on the system size $N$. We {then}
{analyze carefully finite size effects}
to extract typical behaviors in the thermodynamic limit
$N \to \infty$.}

\subsection{Jamming-point}
\label{sec:Jpoint}
To determine the J-point of our model, we follow the standard procedure as in~\cite{o2003jamming, baity2017emergent}, by applying  
 ARC (rapid quench) to random initial configurations ($\varphi_{\rm eq}=0$), and extrapolating the jamming density in the thermodynamical limit  $N \to \infty$ from a finite-size scaling analysis. 
{We {use} $2000$ samples of initial configurations at $\varphi_{\rm eq}=0$  in the following analysis.}

We denote by $f_{\rm IJ} (\varphi_{\rm j}, N, \varphi_{\rm eq})$ the fraction of isotropically jammed realizations at density $\varphi_{\rm j}$, for given $N$ and $\varphi_{\rm eq}$ (for the sake of generality, $\varphi_{\rm eq}$ is expressed as a parameter).
{As shown in Fig.~\ref{fig:finite_size_athermal}(a) (for the case of $\varphi_{\rm eq} = 0$), the fraction shows a finite size effect.}
We  fit $f_{\rm IJ}(\varphi_{\rm j}, N, \varphi_{\rm eq})$ to the form, 
\beq
f_{\rm IJ}(\varphi_{\rm j}, N, \varphi_{\rm eq}) = \frac{1}{2} + \frac{1}{2}\erf \left\{\left[\varphi_{\rm j} - \varphi_{\rm IJ}(N, \varphi_{\rm eq})\right]/w_{\rm IJ}(N,\varphi_{\rm eq}) \right\},
\label{eq:fij}
\eeq 
where $\erf(x)$ is the error function, and $\varphi_{\rm IJ}(N, \varphi_{\rm eq})$ and $\omega_{\rm IJ}(N, \varphi_{\rm eq})$ are fitting parameters.
{The fitting is  shown in Fig.~\ref{fig:finite_size_athermal}(a) by solid lines.}

The jamming density $\varphi_{\rm IJ}(N, \varphi_{\rm eq})$ (see  Fig.~\ref{fig:finite_size_athermal}(c) and Fig.~\ref{fig:scaling_phij}(a)),
and the width $w_{\rm IJ}(N,\varphi_{\rm eq})$ (see  Fig.~\ref{fig:scaling_phij}(d)) show finite size effects.
We then fit $\varphi_{\rm IJ} (N, \varphi_{\rm eq})$ and $w_{\rm IJ}(N,\varphi_{\rm eq})$ to the finite-size scaling forms
\beq
 \varphi_{\rm IJ}(N, \varphi_{\rm eq}) = 
\varphi_{\rm IJ}^{\infty} (\varphi_{\rm eq}) - a
N^{-\mu_{\rm IJ}},
\label{eq:scaling_phiij}
\eeq
and
{
\beq
w_{\rm IJ}(N,\varphi_{\rm eq})=b N^{-\omega_{\rm IJ}}
\label{eq:scaling_wij}
\eeq
where $\varphi_{\rm IJ}^{\infty} (\varphi_{\rm eq})$, $a$ (which depends on $\varphi_{\rm eq}$), $\mu_{\rm IJ}$,
$b$ (which depends on $\varphi_{\rm eq}$) and $w_{\rm IJ}$ are fitting parameters.
The fittings are shown by solid lines in Fig.~\ref{fig:finite_size_athermal}(c), Fig.~\ref{fig:scaling_phij}(a) and  (d), and
the values of  $\varphi_{\rm IJ}^{\infty} (\varphi_{\rm eq})$, $\mu_{\rm IJ}$ and $\omega_{\rm IJ}$ are listed in Table~\ref{table:phi_mu}.}
 
The J-point density for our model is $\varphi_{\rm J}  \equiv \varphi_{\rm IJ}^{\infty} (\varphi_{\rm eq}=0) = 0.655(1)$,  which is consistent with the value reported in Ref.~\cite{ozawa2017exploring}.
 In the thermodynamic limit, {because $\omega_{\rm IJ} > 0$,} $f_{\rm IJ}(\varphi_{\rm j}, N, \varphi_{\rm eq})$ becomes a step function of $\varphi_{\rm j}$ which jumps at $\varphi_{\rm j} = \varphi_{\rm IJ}^{\infty} (\varphi_{\rm eq})$.
 {Our results $\mu_{\rm IJ} = 0.49(1)$ and  $\omega_{\rm IJ}=0.45(1)$ are compatible with the values  $\mu_{\rm IJ} = 0.47(5)$ and  $\omega_{\rm IJ}  = 0.55(3)$ reported in \cite{o2003jamming}} for monodisperse spheres.
Note that the exponent $\mu_{\rm IJ}$ is related to the correlation length exponent $\nu$ in~\cite{o2003jamming} by $\nu = 1/(d\mu_{\rm IJ})$, 
where $d=3$.

\begin{figure}[h!]
\centerline{\includegraphics[width=0.6\columnwidth]{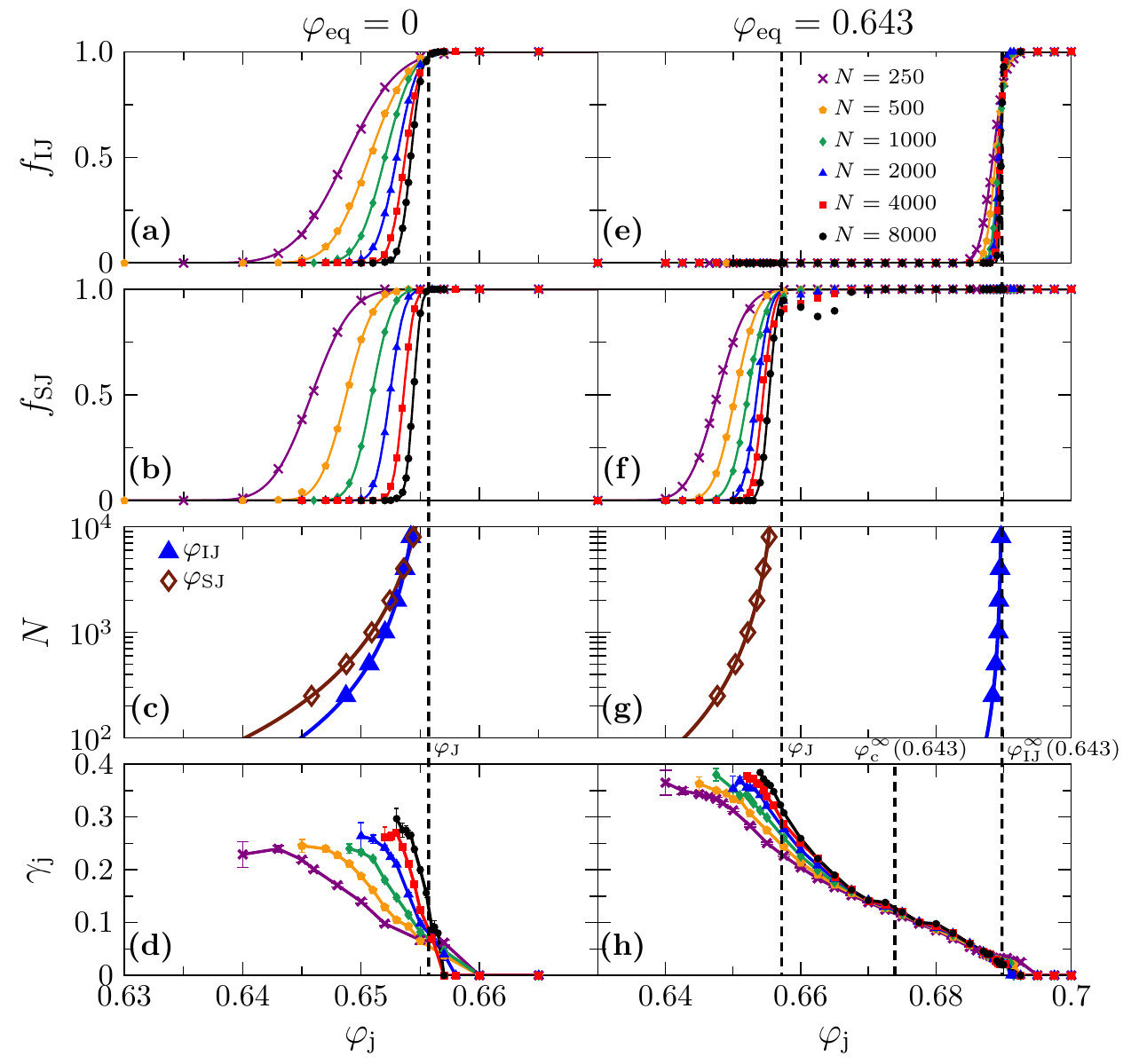}}
\caption{{System-size dependence of isotropic {and} shear jamming in the athermal {protocols}.} { (a-d)} Data for which the initial states are random configurations, $\varphi_{\rm eq}  = 0$.
  {(a)} The fraction of  isotropic jamming $f_{\rm IJ}(\varphi_{\rm j}, N, \varphi_{\rm eq}=0)$ and {(b)} the fraction of shear jamming  $f_{\rm SJ}(\varphi_{\rm j}, N, \varphi_{\rm eq}=0)$
  are plotted as functions of $\varphi_{\rm j}$, for a few different $N$ (points).
  The data points are fitted to Eqs.~(\ref{eq:fij}) and~(\ref{eq:fsj})  (lines).
{(c)} Fitting  $\varphi_{\rm IJ}(N, \varphi_{\rm eq}=0)$ and $\varphi_{\rm SJ}(N, \varphi_{\rm eq}=0)$ to the scaling forms  Eq.~(\ref{eq:scaling_phiij}) and ~(\ref{eq:scaling_phisj}) (see Table~\ref{table:phi_mu} for the fitting parameters) shows that $\varphi_{\rm IJ}^\infty(\varphi_{\rm eq}=0) \simeq \varphi_{\rm SJ}^\infty(\varphi_{\rm eq}=0)$ within the numerical accuracy
({see  Fig.~\ref{fig:scaling_phij}} for the log-log plots).
The vertical dashed line represents $\varphi_{\rm J} =  0.655(1)$.
{(d)} The shear jamming strain $\gamma_{\rm j}$ is plotted as a function of $\varphi_{\rm j}$ for a few different $N$.
{ (e-h)} Same as { (a-d)} but for $\varphi_{\rm eq} =0.643$.
The vertical dashed lines represent $\varphi_{\rm J} = 0.655(1)$, $\varphi_{\rm c}^\infty(\varphi_{\rm eq} = 0.643) = 0.674(1)$, and $\varphi_{\rm IJ}^{\infty}(\varphi_{\rm eq} = 0.643) = 0.690(1)$ (see Table~\ref{table:phi_mu}).
The data for the same $N$ are represented by the same color in (a-b, d-f, h). 
}
\label{fig:finite_size_athermal}
\end{figure}

\begin{figure}[h]
\centerline{\includegraphics[width=0.8\columnwidth]{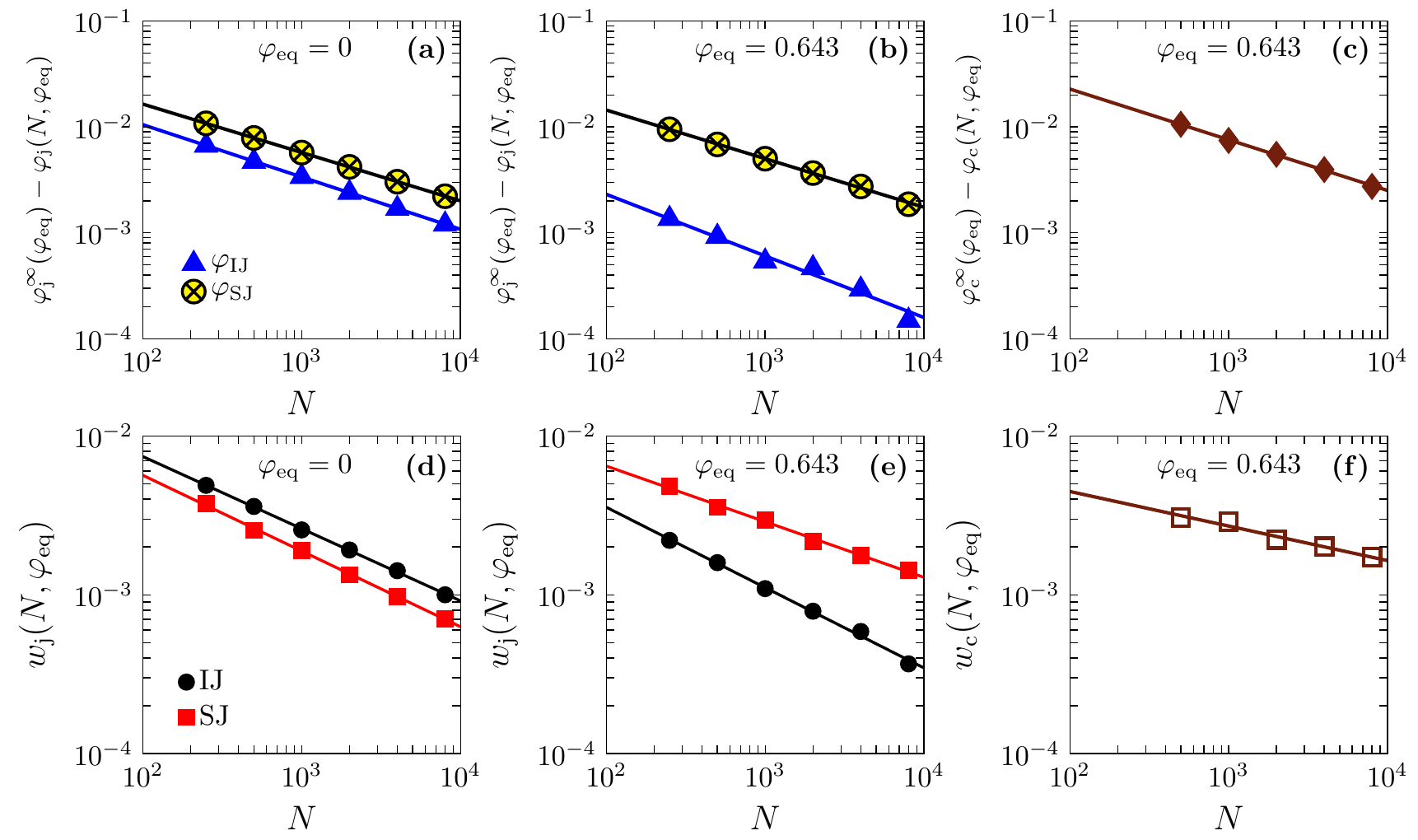}} 
\caption{Finite-size scaling.
 The isotropic and shear jamming  densities obtained by the athermal {protocols}
  are fitted according to the scaling forms Eqs.~(\ref{eq:scaling_phiij}) and  ~(\ref{eq:scaling_phisj})  for 
  {(a)} $\varphi_{\rm eq}=0$  {and}   { (b) } $\varphi_{\rm eq}=0.643$;
their fluctuations are fitted according to Eqs.~(\ref{eq:scaling_wij}) and  ~(\ref{eq:scaling_wsj})  for 
  {(d)} $\varphi_{\rm eq}=0$ {and}    { (e) } $\varphi_{\rm eq}=0.643$.
  (c) The yielding-jamming separation density $\varphi_{\rm c}$ in thermal HSs and (f) its fluctuation are fitted according to Eqs.~(\ref{eq:scaling_phic}) and (\ref{eq:scaling_wc}) for $\varphi_{\rm eq} = 0.643$.
}
\label{fig:scaling_phij}
\end{figure}

\begin{table}[h]
\centering
  \caption{{Numerical values of densities $\varphi_{\rm IJ}^{\infty}$,  $\varphi_{\rm SJ}^{\infty}$, $\varphi_{\rm c}^\infty$, and of exponents $\mu_{\rm IJ}$, $\mu_{\rm SJ}$,  $ \mu_{\rm c}$, $\omega_{\rm IJ}$, $\omega_{\rm SJ}$,  $ \omega_{\rm c}$
      for
      two different $\varphi_{\rm eq}$. }}
\begin{tabular}{ c | c c c |c c c |c c c }
\hline
\hline
$\varphi_{\rm eq}$ & $\varphi_{\rm IJ}^{\infty}$ & $\varphi_{\rm SJ}^{\infty}$ & $\varphi_{\rm c}^\infty$ &  $\mu_{\rm IJ}$ & $\mu_{\rm SJ}$ & $\mu_{\rm c}$
&  $\omega_{\rm IJ}$ & $\omega_{\rm SJ}$ & $\omega_{\rm c}$
\\
\hline
 0  & 0.655(1) & 0.657(1)   & - & 0.49(1) &  0.46(1) & - & 0.45(1) & 0.48(1) & - \\
0.643& 0.690(1) & 0.657(1)  & 0.674(1) & 0.6(1) & 0.46(3)  & 0.48(5) & 0.51(2) & 0.35(1) & 0.22(2) \\
\hline
\hline  
\end{tabular}
\label{table:phi_mu}
\end{table}

\subsection{{Jamming-line: the athermal case}}
\label{sec:Jline_athermal}
As demonstrated in Refs.~\cite{ozawa2012jamming, ozawa2017exploring}, the J-point at $\varphi_{\rm J}$
can be extended into a J-line Eq.~(1) 
by replacing 
the initial random configurations ($\varphi_{\rm eq}=0$) with dense equilibrium configurations  ($\varphi_{\rm eq} > 0$), 
before applying ARC.
Figure~\ref{fig:Jline_SS} shows  the dependence of $\varphi_{\rm IJ}(N, \varphi_{\rm eq})$ on $\varphi_{\rm eq}$, for two different $N$. 
Below the onset density $\varphi_{\rm onset} \approx 0.56$, the isotropic jamming density $\varphi_{\rm IJ} $ is nearly a constant; above $\varphi_{\rm onset}$, it increases monotonically with $\varphi_{\rm eq}$~\cite{ozawa2012jamming, ozawa2017exploring}. 
For  $\varphi_{\rm eq} \in [0, \varphi_{\rm eq}^{\rm max}]$, we generate jammed packings on a J-line covering a  range of densities, $\varphi_{\rm IJ} \in [\varphi_{\rm IJ}^{\rm min, ath}, \varphi_{\rm IJ} ^{\rm max, ath}]$. In the thermodynamical limit, the lower bound  is set by the J-point density, $\varphi_{\rm IJ}^{\rm min, ath} = \varphi_{\rm J}$, as discussed in {Sec.}~\ref{sec:Jpoint}. The maximum jamming density $\varphi_{\rm IJ} ^{\rm max, ath} = \varphi_{\rm IJ}( \varphi_{\rm eq}^{\rm max}) $ depends on the maximum equilibrium density that we are able to reach in the protocol of preparing initial configurations. In this study,  we obtain $\varphi_{\rm eq}^{\rm max} = 0.655$ for $N=1000$, and $\varphi_{\rm eq}^{\rm max} = 0.643$ for $N=8000$, corresponding to $\varphi_{\rm IJ}^{\rm max, ath} = 0.695$ and $\varphi_{\rm IJ}^{\rm max, ath} = 0.690$ respectively~\cite{berthier2016growing}. 
 
Not only the J-point ($\varphi_{\rm eq} < \varphi_{\rm onset}$), but also any other point ($\varphi_{\rm eq} > \varphi_{\rm onset}$) on the J-line satisfies the finite-size formulas Eqs.~(\ref{eq:fij}),~(\ref{eq:scaling_phiij}) and ~(\ref{eq:scaling_wij}).
In Fig.~\ref{fig:finite_size_athermal}(e) and Fig.~\ref{fig:scaling_phij}(b) and (e),
we show an example with $\varphi_{\rm eq} = 0.643$.
Our results are compatible with
$\mu_{\rm IJ} \approx w_{\rm IJ} \approx 0.5$~\cite{baity2017emergent},
independent of $\varphi_{\rm eq}$  (see Table~\ref{table:phi_mu}).    
{In the analysis, $500$ samples of initial configurations at $\varphi_{\rm eq}=0.643$ are used.}



\begin{figure}[h]
\centerline{\includegraphics[width=0.4\columnwidth]{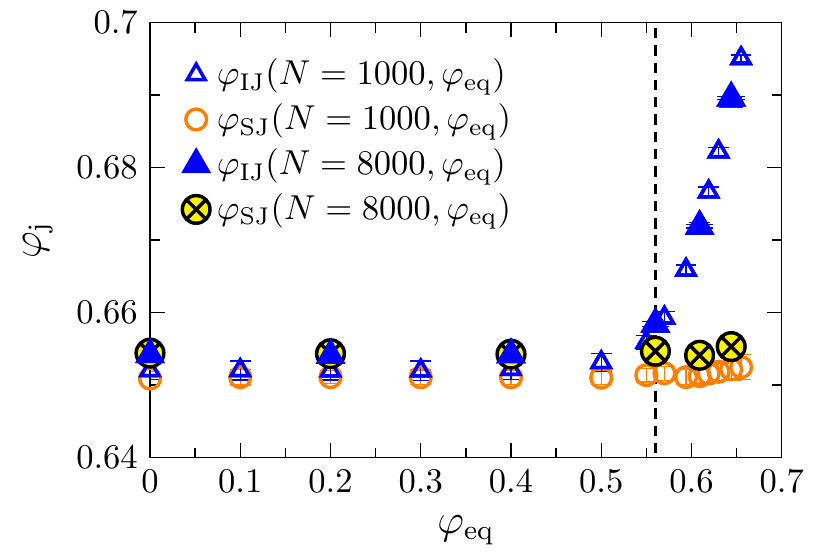}} 
\caption{ Isotropic jamming density $\varphi_{\rm IJ}$  and shear jamming density $\varphi_{\rm SJ}$ as functions of
  $\varphi_{\rm eq}$, for $N=1000$ and $8000$, obtained by the athermal protocol. The onset density $\varphi_{\rm onset} \approx 0.56$ is marked by the vertical dashed line.
{{Date are averaged over} $200-2000$ independent samples for each $\varphi_{\rm eq}$ here.}
}
\label{fig:Jline_SS}
\end{figure}

\subsection{Jamming-plane: {the athermal case}}
\label{sec:Jplane_ath}
The J-plane is extended from the J-line by including shear jammed states. 
Starting from the remaining $1-f_{\rm IJ}(\varphi_{\rm j}, N, \varphi_{\rm eq})$ fraction of unjammed configurations after the 
ARC procedure, we apply AQS that stops either when the system jams or the strain exceeds $\gamma_{\rm max}$.
We denote by $f_{\rm SJ} (\varphi_{\rm j}, N, \varphi_{\rm eq})$ the fraction of configurations that jam under shear,
and by $\gamma_{\rm j} (\varphi_{\rm j}, N, \varphi_{\rm eq})$ the average jamming strain.
In total, there are a fraction of  $(1-f_{\rm IJ})(1-f_{\rm SJ})$ samples do not jam even  
at $\gamma=\gamma_{\rm max}$.
Similarly to the isotropic jamming case \eq{eq:fij}, we fit the fraction of shear jamming  $f_{\rm SJ}(\varphi_{\rm j}, N, \varphi_{\rm eq})$  to the form 
\beq
f_{\rm SJ}(\varphi_{\rm j}, N, \varphi_{\rm eq})
= \frac{1}{2}
+ \frac{1}{2}\erf \left\{\left[\varphi_{\rm j} - \varphi_{\rm SJ}(N, \varphi_{\rm eq})\right]/w_{\rm SJ}(N,\varphi_{\rm eq})\right \},
\label{eq:fsj}
\eeq 
and use the finite-size scalings
\beq
 \varphi_{\rm SJ}(N, \varphi_{\rm eq}) = \varphi_{\rm SJ}^{\infty} (\varphi_{\rm eq}) - a N^{-\mu_{\rm SJ}}
\label{eq:scaling_phisj}
\eeq
and
\beq
w_{\rm SJ}(N,\varphi_{\rm eq})=b  N^{-\omega_{\rm SJ}}
\label{eq:scaling_wsj}
\eeq
to estimate $\varphi_{\rm SJ}^{\infty}(\varphi_{\rm eq})$, $\mu_{\rm SJ}$ and  $\omega_{\rm SJ}$ (see Table~\ref{table:phi_mu}  for the values).

\subsubsection{Shear jamming for $\varphi_{\rm eq}=0$}
  
Let us first discuss the case of $\varphi_{\rm eq}=0$ for shear jamming. 
Note that as long as $\varphi_{\rm eq} < \varphi_{\rm onset}$, the behavior of isotropic jamming (Fig.~\ref{fig:Jline_SS}), as well as {that of } shear jamming, is independent of $\varphi_{\rm eq}$.
The fittings of Eqs.~(\ref{eq:fsj}),~(\ref{eq:scaling_phisj}) and
~(\ref{eq:scaling_wsj}) are included in Figs.~\ref{fig:finite_size_athermal}(b),~\ref{fig:scaling_phij}(a) and~(d).
According to the values listed in Table~\ref{table:phi_mu},
we find that  $\varphi_{\rm IJ}^{\infty} (\varphi_{\rm eq}=0) \simeq \varphi_{\rm SJ}^{\infty} (\varphi_{\rm eq}=0) \simeq \varphi_{\rm J}$  within the numerical error. 
Therefore, isotropic jamming and shear jamming occur at the same density in the thermodynamical limit, which is consistent with  Ref.~\cite{baity2017emergent}.
However, we do not exclude the possibility that there is a small difference, which is in the order of 0.001, between the isotropic jamming density $\varphi_{\rm IJ}^{\infty} (\varphi_{\rm eq}=0)$ and the shear jamming density $\varphi_{\rm SJ}^{\infty} (\varphi_{\rm eq}=0)$, in the thermodynamical limit~{\cite{kawasaki2020shear}}.
The scaling exponents {obtained from} 
the finite-size {scalings} Eqs.~(\ref{eq:scaling_phiij}),~(\ref{eq:scaling_wij}), ~(\ref{eq:scaling_phisj})  and  ~(\ref{eq:scaling_wsj}) are also approximately identical, $\mu_{\rm IJ} \approx \mu_{\rm SJ} \approx \omega_{\rm IJ} \approx \omega_{\rm SJ} \approx 0.5$~\cite{baity2017emergent}. 

Since $\varphi_{\rm IJ}^{\infty} (\varphi_{\rm eq}=0) =  \varphi_{\rm SJ}^{\infty}(\varphi_{\rm eq}=0)$, the SJ-line Eq.~(2) 
should be vertical in the thermodynamical limit~\cite{baity2017emergent, bertrand2016protocol}, which is consistent with the trend seen in Fig.~\ref{fig:finite_size_athermal}(d). In fact, we can show that the SJ-lines $\gamma_{\rm j} (\varphi_{\rm j}, N,\varphi_{\rm eq}=0)$ collapse onto a master curve $\gamma_{\rm j} \left[ (\varphi_{\rm j}  - \varphi_{\rm SJ}^{\infty})N^\mu, \varphi_{\rm eq}=0 \right]$, with  $\mu = 1/2$ (see Fig.~\ref{fig:strain_rapid_quench}), as suggested in Ref.~\cite{baity2017emergent}.
 The dependence of the SJ-line on the protocol parameter $\gamma_{\rm max}$ further indicates that
 the SJ-line 
   extends to infinite strain    $\gamma_{\rm j} \to \infty$ in the limit $\gamma_{\rm max} \to \infty$ (Fig.~\ref{fig:strain_rapid_quench} inset).

   \subsubsection{Shear jamming for $\varphi_{\rm eq} > \varphi_{\rm onset}$}
 
   We next discuss the case of $\varphi_{\rm eq}= 0.643$ (Fig.~\ref{fig:finite_size_athermal}(e)-(h)), as an example for $\varphi_{\rm eq} > \varphi_{\rm onset}$.
   In contrast to the previous case ($\varphi_{\rm eq}= 0$),   the shear jamming density is unambiguously lower than the isotropic jamming density in the thermodynamic limit, $\varphi_{\rm SJ}^{ \infty} (\varphi_{\rm eq} = 0.643) < \varphi_{\rm IJ}^{ \infty} (\varphi_{\rm eq} = 0.643)$  (see Table~\ref{table:phi_mu}  for the values), and the SJ-line  $\gamma_{\rm j} (\varphi_{\rm j}, N\to\infty, \varphi_{\rm eq}=0.643)$ does not tend to be vertical  {in that limit} (Fig.~\ref{fig:finite_size_athermal}(h)). The exponents $\mu_{\rm IJ} \approx \mu_{\rm SJ} \approx 0.5$ are independent of $\varphi_{\rm eq}$ within the numerical accuracy, {while {$\omega_{\rm SJ} = 0.35(1)$} is slightly smaller than {$\omega_{\rm IJ}=0.51(2)$}} (see Fig.~\ref{fig:finite_size_athermal}(f), Fig.~\ref{fig:scaling_phij}(b, e) and  Table~\ref{table:phi_mu}). 

We then extend our analysis more systematically to general values of $\varphi_{\rm eq}$.
As shown in Fig.~\ref{fig:Jline_SS},
 $\varphi_{\rm SJ}(N,\varphi_{\rm eq})$ is nearly a constant as a function of $\varphi_{\rm eq}$. 
Figure~\ref{fig:Jline_SS} also shows that $\varphi_{\rm IJ} \simeq \varphi_{\rm SJ}$ when $\varphi_{\rm eq}$ is below $\varphi_{\rm onset} \approx 0.56$, and {that} 
$\varphi_{\rm IJ} > \varphi_{\rm SJ}$ above. These results indicate that, for $\varphi_{\rm eq} > \varphi_{\rm onset}$, shear jamming could occur over a range of densities $\varphi_{\rm j} \in [\varphi_{\rm SJ}, \varphi_{\rm IJ}]$. 
{It is this newly}
discovered region {that} extends the phase space of jamming to a J-plane. 
In Fig.~\ref{fig:Jplane_athermal}, we plot 
the SJ-lines Eq.~(2), 
for a few different  $\varphi_{\rm eq}$ to represent the J-plane.


\begin{figure}[h]
\centerline{\includegraphics[width=0.4\columnwidth]{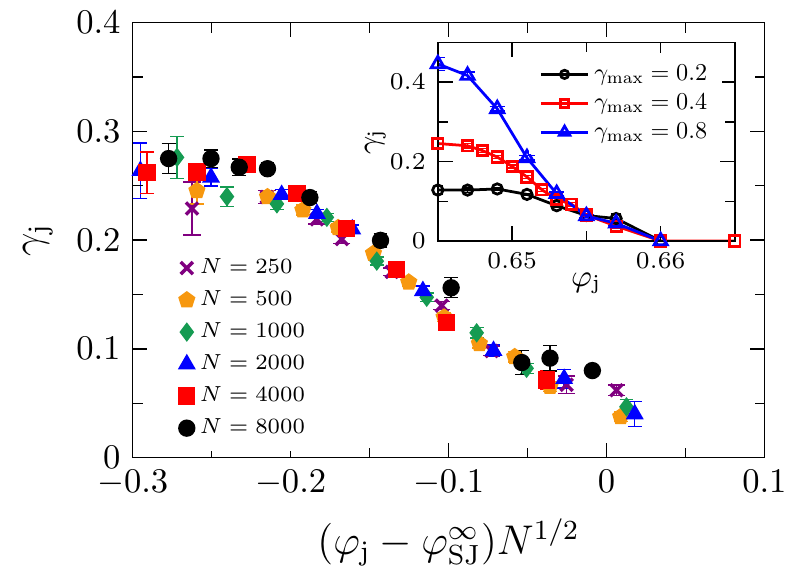}} 
\caption{Collapse of SJ-lines obtained by the athermal protocols, for $\varphi_{\rm eq}=0$ and a few different $N$, where $\varphi_{\rm SJ}^{\infty} = 0.657$ is used (see Table~\ref{table:phi_mu}).  (inset) The SJ-line, for $\varphi_{\rm eq}=0$ and $N=500$, becomes steeper with increasing  $\gamma_{\rm max}$.}
\label{fig:strain_rapid_quench}
\end{figure}

\begin{figure}[h]
\centerline{\includegraphics[width=0.4\columnwidth]{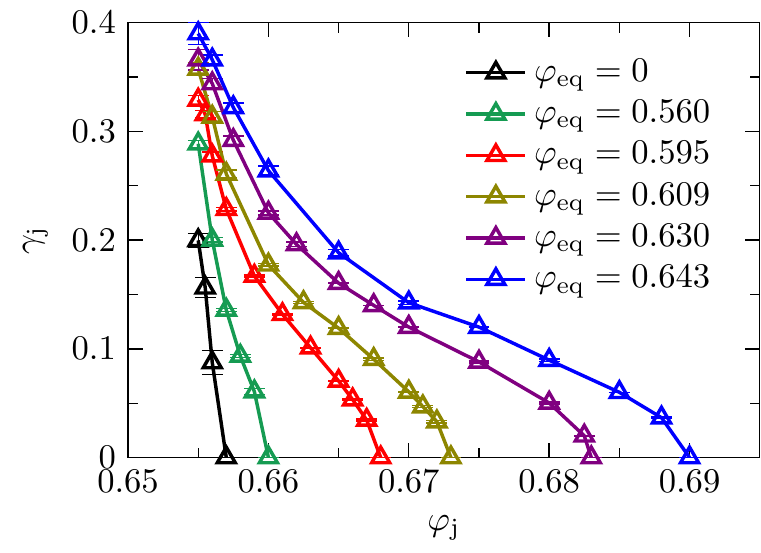}} 
\caption{J-plane obtained by the athermal {protocols} ($N=8000$), represented by typical SJ-lines $\gamma_{\rm j} (\varphi_{\rm j}; \varphi_{\rm eq})$ for  a few different  $\varphi_{\rm eq}$ (we only show the part for $\varphi_{\rm j} > \varphi_{\rm SJ}$, which means that $f_{\rm SJ} > 0.5$).}
\label{fig:Jplane_athermal}
\end{figure}


\section{Exploring the jamming-plane using {thermal protocols}}
\label{sec:thermal}
{Similar  to the athermal case, we apply TC to obtain isotropically jammed packings, and TC+TQS to obtain shear jammed packings, 
starting from a large number of independent samples of initial states generated at $\{\varphi_{\rm eq},0\}$.
%
}

\subsection{Minimum isotropic jamming density}
\label{sec:min_phiJ_th}
The minimum isotropic jamming density $\varphi_{\rm IJ}^{\rm min, th}$ is obtained, in principle, in the limits of $\Gamma \to \infty$ and $\varphi_{\rm eq} \to 0$, because  $\varphi_{\rm IJ}(\varphi_{\rm eq}, \Gamma)$ decreases with  increasing $\Gamma$ (Fig.~\ref{fig:threshold_point}(a)) or  decreasing $\varphi_{\rm eq}$ (Fig.~\ref{fig:Jline_HS}).
However, Fig.~\ref{fig:threshold_point}(b) shows that the packings generated by the {TC} protocol are isostatic, i.e.
{$z_{\rm j}=6$}, only when $\Gamma 
\leq 3 \times 10^{-4}$. Based on that, we determine $\varphi_{\rm IJ}^{\rm min, th} = \varphi_{\rm IJ}(\varphi_{\rm eq}=0, \Gamma = 3 \times 10^{-4}) = 0.665$,
{which is slightly larger than
  the J-point density $\varphi_{\rm J}=0.655$}.
We restrict our  simulations to small compression rates $\Gamma \leq 3 \times 10^{-4}$, that is, only  isostatic packings are considered.


\subsection{State-following  jamming density}
\label{sec:SF}
Figure~\ref{fig:Jline_HS} shows that, when $\varphi_{\rm eq}$ is above a certain density  $\varphi_{\rm SF} \approx 0.60$,
the function $\varphi_{\rm IJ}(\varphi_{\rm eq} )$ becomes linear and independent of the compression rate $\Gamma$. The data obtained from the ARC, which corresponds to the limit of infinitely rapid quench ($\Gamma \to \infty$), also collapses onto the same liner function  for $\varphi_{\rm eq}>\varphi_{\rm SF}$. 
We define the state-following jamming density  $\varphi_{\rm J}^{\rm SF} = \varphi_{\rm IJ}(\varphi_{\rm SF}) \approx 0.67$  
 as  the minimum density of isotropically jammed states that can be obtained independent of the compression rate $\Gamma$. 
Above {$\varphi_{\rm eq} = \varphi_{\rm SF}$}, 
the final state after quench 
can be unambiguously  mapped 
onto the initial state, which is the reason why the quench dynamics are called  ``state-following"~\cite{franz1995recipes}. We point out that, for sufficiently small $\Gamma$, activated dynamics will play a role. However, such small $\Gamma$ cannot be reached in our current simulations, and will not be considered. 
In {Sec.}~\ref{sec:reversibility}, we will discuss the connection between the state-following  quench dynamics and the reversibility of jamming under shear.



\begin{figure}[h]
\centerline{\includegraphics[width=0.4\columnwidth]{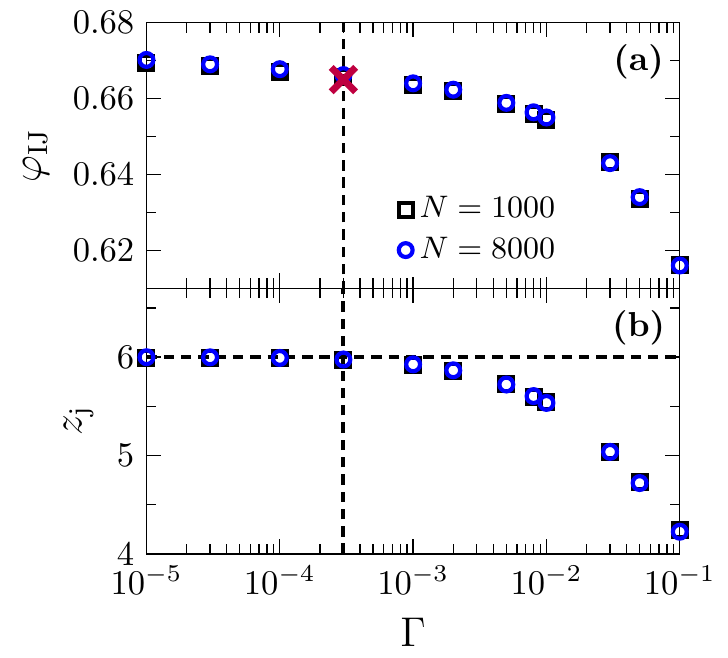}} 
\caption { (a)  Isotropic jamming density $\varphi_{\rm IJ}$ and (b) average coordination number $z_{\rm j}$ (without rattlers) as functions of compression rate $\Gamma$, for $\varphi_{\rm eq} = 0$ and two different $N$, obtained by the  {TC} protocol.
The vertical dashed line represents $\Gamma = 3 \times 10^{-4}$, and the cross represents $\varphi_{\rm IJ}^{\rm min, th} = 0.665$.
No appreciable system-size dependence is observed. 
}
\label{fig:threshold_point}
\end{figure}

\subsection{Jamming-line: the thermal case}
\label{sec:Jline-thermal}
The J-line Eq. (1),
which is here generalized to 
\beq
\varphi_{\rm IJ}=\varphi_{\rm IJ}(\varphi_{\rm eq}, \Gamma),
\eeq
is obtained by varying the compression rate $\Gamma$ and the initial density $\varphi_{\rm eq}$ (see Fig.~\ref{fig:Jline_HS}). 
It covers a range of jamming densities,  $\varphi_{\rm IJ}(\varphi_{\rm eq}, \Gamma) \in [\varphi_{\rm IJ}^{\rm min, th}, \varphi_{\rm IJ}^{\rm max, th}]$.
For a given $\Gamma$, the isotropic jamming density  $\varphi_{\rm IJ}$ decreases with decreasing $\varphi_{\rm eq}$, and becomes independent of $\varphi_{\rm eq}$ below $\varphi_{\rm onset}$. 
The minimum value $ \varphi_{\rm IJ}(\varphi_{\rm eq} < \varphi_{\rm onset}, \Gamma)$ itself 
 decreases with increasing $\Gamma$, and converges to $\varphi_{\rm IJ}^{\rm min, th} = 0.665$ as $\Gamma \to 3 \times 10^{-4}$ (see also {Sec.}~\ref{sec:min_phiJ_th}).
The upper bound $\varphi_{\rm IJ}^{\rm max, th}$, on the other hand, only depends on $\varphi_{\rm eq}^{\rm max}$ as in the athermal protocol, and 
is independent of $\Gamma$.
Because $\varphi_{\rm IJ}^{\rm min, th} > \varphi_{\rm J}$, 
it is impossible to reach the J-point in the thermal protocol.
It also means that the thermal protocol explores a narrower part of the J-line compared to the athermal protocol.
We have checked that the finite-size effect of $\varphi_{\rm IJ}$ in the thermal protocol is negligible, see Fig.~\ref{fig:threshold_point}(a) for the case of $\varphi_{\rm eq}=0$, and Fig.~\ref{fig:phic}(b) (blue triangles) for the case of   $\varphi_{\rm eq} = 0.643$.


\begin{figure}[h]
  \centerline{\includegraphics[width=0.4\columnwidth]{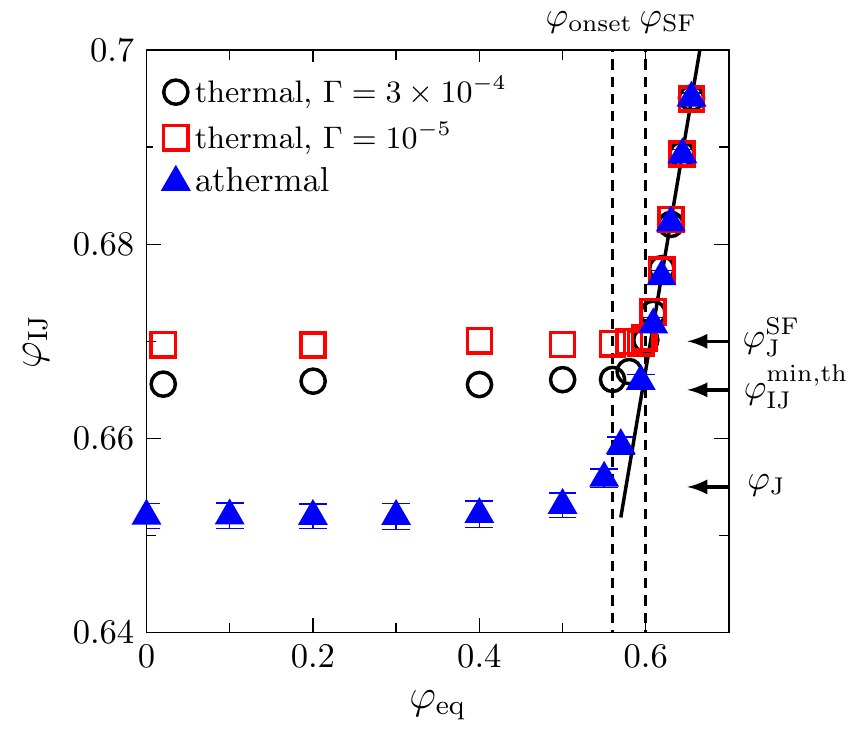}}
\caption{ Isotropic jamming density $\varphi_{\rm IJ}$ as a function of
$\varphi_{\rm eq}$, for $N=1000$ and two different compression rates $\Gamma$, obtained by the {TC} protocol ({compare with Fig.~\ref{fig:Jline_SS} for the athermal case}). 
The  
$N=1000$ athermal data from Fig.~\ref{fig:Jline_SS} is also plotted. 
The vertical lines mark $\varphi_{\rm onset} \approx 0.56$ 
and $\varphi_{\rm SF} \approx 0.60$.  The horizontal arrows mark jamming densities $\varphi_{\rm J}  = 0.655(1)$, $\varphi_{\rm IJ}^{\rm min, th} =  0.665$, and $\varphi_{\rm J}^{\rm SF} \approx 0.67$. The solid line represents the function $\varphi_{\rm IJ} (\varphi_{\rm eq})= 0.51 \varphi_{\rm eq} + 0.36$ obtained from linear fitting 
to the data in the regime $\varphi_{\rm eq} \geq \varphi_{\rm SF}.$
}
\label{fig:Jline_HS}
\end{figure}

\subsection{Jamming-plane: the thermal case}
\label{sec:Jplane_th}
Different from the athermal case,  thermal HSs do not only display the behavior of jamming 
under constant volume shear. 
A HS glass  quenched from $\varphi_{\rm eq}$ {\it yields} if the shear deformation is applied at $\varphi < \varphi_{\rm c}(\varphi_{\rm eq})$, and jams  if $\varphi > \varphi_{\rm c}(\varphi_{\rm eq})$~\cite{urbani2017shear, jin2018stability}, see Fig.~\ref{fig:yielding_jamming}(a). Therefore, shear yielding and shear jamming are separated by the yielding-jamming separation point  at  $\varphi_{\rm c}$~\cite{jin2018stability}.
According to the mean-field theory~\cite{urbani2017shear}, this point is a critical point, which however is not the case in finite dimensions~\cite{jin2018stability}. 
In Fig.~\ref{fig:yielding_jamming}(b), we plot $\varphi_{\rm c}(\varphi_{\rm eq})$ for a few different $\varphi_{\rm eq}$ on the 
 HS phase diagram, which is qualitatively consistent with the mean-field result~\cite{altieri2019mean}.
To avoid confusion we note that Fig.~\ref{fig:yielding_jamming}(a) shows
   the {\it entropic} stress-strain curve of thermal HSs, which must be distinguished from the
{\it mechanical} stress-strain curve of athermal SSs {(Fig.~\ref{fig:stress_controlled})}.   Athermal yielding of SSs happens above jamming.


We next perform a finite-size analysis of thermal shear jamming. 
We denote the fraction of shear jamming by
{$f_{\rm c}(\varphi_{\rm j}, N, \varphi_{\rm eq})$ (and therefore $1-f_{\rm c}(\varphi_{\rm j}, N, \varphi_{\rm eq})$ is the fraction of shear yielding)
and by {$\gamma_{\rm j} (\varphi_{\rm j}, N, \varphi_{\rm eq})$ the average jamming strain}.
As an example, we fit the data of $f_{\rm c}(\varphi_{\rm j}, N, \varphi_{\rm eq} = 0.643)$  to the form (see Fig.~\ref{fig:phic}(a)),}
\beq
f_{\rm c}(\varphi_{\rm j}, N, \varphi_{\rm eq}) = \frac{1}{2} + \frac{1}{2}\erf \left\{\left[\varphi_{\rm j} - \varphi_{\rm c}(N, \varphi_{\rm eq})\right]/w_{\rm c}(N,\varphi_{\rm eq})\right \}.
\label{eq:fsj_HS}
\eeq
We then estimate the values of the asymptotic density $\varphi_{\rm c}^{\infty} (\varphi_{\rm eq})$,
the exponents  $\mu_{\rm c}$ and $\omega_{\rm c}$ (see Table~\ref{table:phi_mu})
using the finite-size scaling forms
{(see Fig.~\ref{fig:phic}(b), Fig.~\ref{fig:scaling_phij}(c) and (f))},
\beq
\varphi_{\rm c}(N, \varphi_{\rm eq}) = \varphi_{\rm c}^{\infty} (\varphi_{\rm eq}) - a  N^{-\mu_{\rm c}},
\label{eq:scaling_phic}
\eeq
and
\beq
w_{\rm c}(N,\varphi_{\rm eq})=b  N^{-\omega_{\rm c}}.
\label{eq:scaling_wc}
\eeq
The data in Table~\ref{table:phi_mu} show that $\varphi_{\rm SJ}^\infty < \varphi_{\rm c}^\infty < \varphi_{\rm IJ}^\infty$. The exponents of $\mu$, for both athermal and thermal cases, are universal within the numerical accuracy, $\mu_{\rm IJ} \approx \mu_{\rm SJ} \approx  \mu_{\rm c} \approx 0.5$, {while $\omega_{\rm c}$ is clearly smaller than $0.5$.}
\red{The result $\mu \approx 0.5$ suggests  that the value of  correlation length critical exponent is $\nu = 1/(d \mu) \approx 0.7$.
By taking into account the next-order scaling corrections, Ref.~\cite{vaagberg2011finite} obtained $\nu \approx 1$ for jammed packings generated  from both rapid quench ($\varphi_{\rm eq} = 0$) and quasi-static shearing.
In principle, scaling corrections can be also considered for the case of  $\varphi_{\rm eq} > 0$, but the present 
 data are not sufficient for that purpose. 
}
The SJ-lines  Eq.~(2), 
obtained by the thermal protocol, have negligible finite-size effects (Fig.~\ref{fig:phic}(c)).
To represent the J-plane,  in Fig.~\ref{fig:Jplane_thermal} we plot the SJ-lines for a few different  $\varphi_{\rm eq}$ obtained by the thermal protocol.

We compare the J-plane obtained by the athermal 
(see Fig.~\ref{fig:Jplane_athermal}) and  the thermal (see Fig.~\ref{fig:Jplane_thermal})  protocols in Fig.~2(a).
  The SJ-lines obtained by the two protocols match,
  but the thermal protocol explores a smaller region of the  J-plane than the athermal protocol.
We  explain in {Sec.}~\ref{sec:reversibility} the reason for this difference. 



\begin{figure}[h!]
\centerline{\includegraphics[width=0.7\columnwidth]{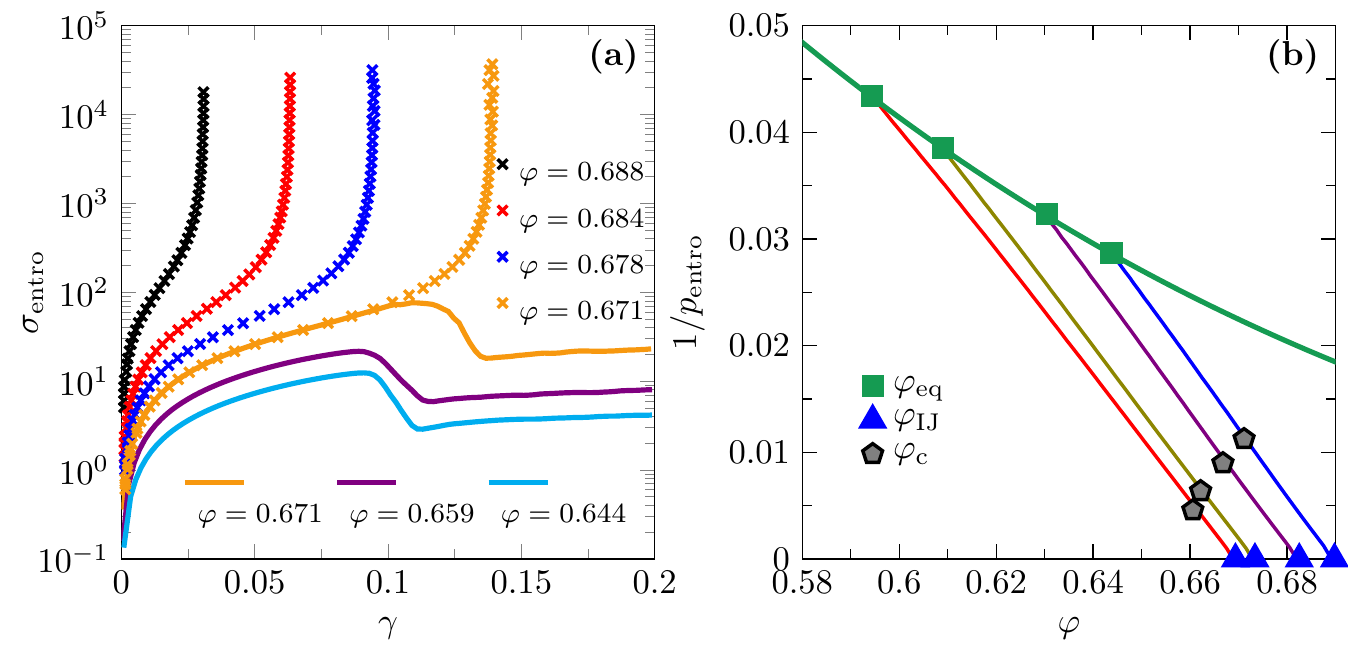}} 
\caption{ Shear yielding and shear jamming in thermal HSs. {(a)} {Entropic} stress-strain curves at a few different $\varphi$, for $\varphi_{\rm eq} = 0.643$ and $N=8000$.
{(b)} Yielding-jamming separation points $\varphi_{\rm c}(\varphi_{\rm eq})$  on the $1/p_{\rm entro} - \varphi$ phase diagram. The green line represents the liquid EOS $p_{\rm entro}^{\rm L}(\varphi_{\rm eq})$, and the other lines represent the glass EOSs $p_{\rm entro}^{\rm G}(\varphi; \varphi_{\rm eq})$ for $\varphi_{\rm eq} = 0.595, 0.609, 0.630, 0.643$ (from left to right).}
\label{fig:yielding_jamming}
\end{figure}

\begin{figure}[h!]
 \centerline{\includegraphics[width=0.35\columnwidth]{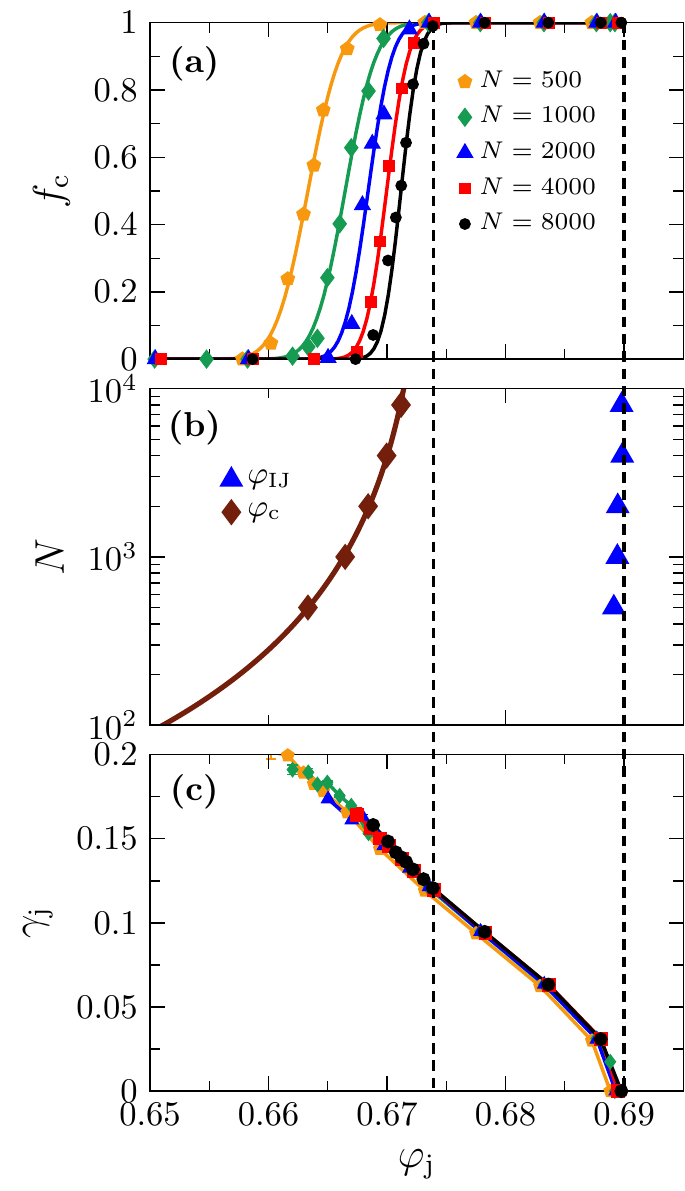}} 
\caption{System-size dependence of jamming in the thermal
{protocols} for $\varphi_{\rm eq} = 0.643$ (compare with Fig.~\ref{fig:finite_size_athermal} for the athermal case). (a) Fraction $f_{\rm c}$ of  shear jamming as a function of $\varphi_{\rm j}$, for a few different $N$. The data points are fitted to Eq.~(\ref{eq:fsj_HS}) (lines).
{(b)} The data of  $\varphi_{\rm c}(N)$ is fitted to Eq.~(\ref{eq:scaling_phic}), see Fig.~\ref{fig:scaling_phij}(c) for the log-log plot.
We do not attempt to fit $\varphi_{\rm IJ}(N)$ to any scaling forms, since the $N$-dependence is negligible {(see also Fig.~\ref{fig:threshold_point}(a))}. 
The asymptotic densities (see Table~\ref{table:phi_mu}), $\varphi_{\rm c}^\infty = 0.674(1)$, and $\varphi_{\rm IJ}^\infty = 0.690(1)$ that is obtained by the athermal protocol for the same $\varphi_{\rm eq}$, are marked by vertical dashed lines. {(c)} The shear jamming strain $\gamma_{\rm j}$ is plotted as a function of $\varphi_{\rm j}$ for a few different $N$, which shows that 
 the system-size dependence is negligible. 
}
\label{fig:phic}
\end{figure}

\begin{figure}[h]
\centerline{\includegraphics[width=0.4\columnwidth]{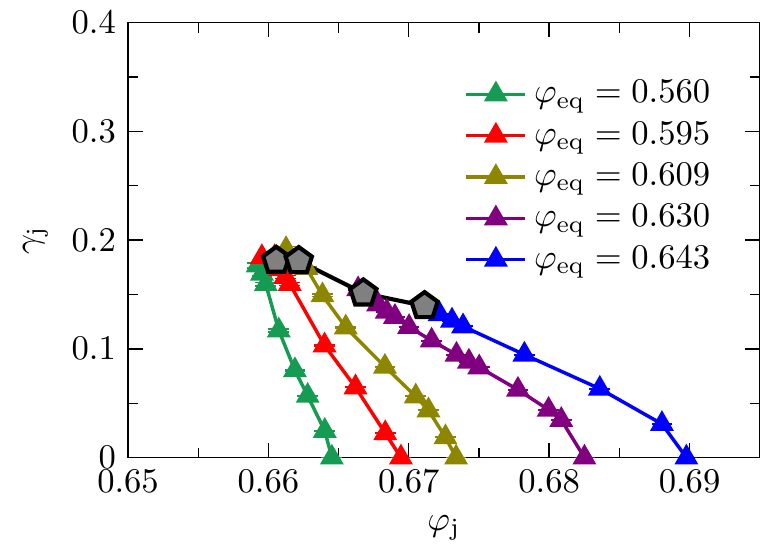}} 
\caption{J-plane obtained by the thermal protocol, represented by typical SJ-lines $\gamma_{\rm j} (\varphi_{\rm j}; \varphi_{\rm eq})$ for a few different  $\varphi_{\rm eq}$ ($N=8000$). 
  These SJ-lines are bounded from above by the yielding-jamming separation line $\{\varphi_{\rm c}(\varphi_{\rm eq}), \gamma_{\rm c}(\varphi_{\rm eq})\}$ (pentagons), where $\gamma_{\rm c}(\varphi_{\rm eq})  = \gamma_{\rm j} (\varphi_{\rm j} = \varphi_{\rm c}; \varphi_{\rm eq})$.
{Compare with Fig.~\ref{fig:Jplane_athermal} for the athermal case.}
}
\label{fig:Jplane_thermal}
\end{figure}

\section{Reversibility}
\label{sec:reversibility}
{We have} discussed how to reach jamming at $\{\varphi_{\rm j}, \gamma_{\rm j} \}$
  starting from HS liquid states at $\{\varphi_{\rm eq} , 0\}$, using either athermal or thermal protocols.
  Here we ask whether such routes to jamming can be reversed.
 We   point out that the reversibility mainly depends on the state variables $\varphi_{\rm j}$ and $\gamma_{\rm j}$, with an intrinsic relationship to $\varphi_{\rm eq}$, rather than on the protocol itself (athermal or thermal). The difference between athermal and thermal protocols lies in the accessibility to irreversible-jamming.
 In the following, we will focus on using the athermal protocols (ARC and AQS) to examine the reversibility.
  In \cite{jin2018stability}, we have found that the routes to shear jamming are reversible under TQS, for the few cases studied, which have $\varphi_{\rm eq} > \varphi_{\rm SF}$ and $\varphi_{\rm j} > \varphi_{\rm c}(\varphi_{\rm eq})$. That observation is consistent with the systematic study presented below using the athermal protocols.

\subsection{Definition}

In {Sec.}~\ref{sec:athermal}, we {have} described how to use the athermal {protocols} to generate a jammed configuration at
$\{ \varphi_{\rm j},\gamma_{\rm j}(\varphi_{\rm j}; \varphi_{\rm eq}) \}$, 
from the initial equilibrium state at
$\{ \varphi_{\rm eq},0 \}$.
The procedure consists of two steps,
  \begin{eqnarray}
{\rm ARC}&:& \qquad \{\varphi_{\rm eq},0\} \to  \{\varphi_{\rm j},0\}, \label{eq:compression_route} \\
{\rm AQS}&:& \qquad  \{\varphi_{\rm j}, 0\} \to \{\varphi_{\rm j}, \gamma_{\rm j}(\varphi_{\rm j}; \varphi_{\rm eq})\}. \label{eq:shear_route}
  \end{eqnarray}
Let us recall that a fraction of $f_{\rm IJ}$
 samples jam at $\{\varphi_{\rm j}, 0 \}$ simply by the ARC procedure 
(\eq{eq:compression_route}) without adding shear.
For the rest of unjammed  samples, 
we further apply AQS (\eq{eq:shear_route}) 
to shear jam them at $\{\varphi_{\rm j},\gamma_{\rm j}(\varphi_{\rm j}; \varphi_{\rm eq})\}$.
Therefore, in principle we can study the reversibility of the two steps, compression jamming and shear jamming, separately.
It turns out that 
each reversible-jamming state, either compression or shear jammed,  
 is uniquely associated with a metastable glass basin quenched from $\{\varphi_{\rm eq},0\}$.

To quantify the reversibility, we apply a single cycle of {compression or} shear,
and measure the relative mean square displacement  (RMSD) $\Delta_{\rm r} = \frac{1}{N} \sum_{i=1}^{N} \left| \vec{r}_i^{\rm after} - \vec{r}_i^{\rm before}  \right|^2$
 between the configuration  $\{\vec{r}_i^{\rm before} \}$ before the cycle and the configuration $\{\vec{r}_i^{\rm after} \}$ after.
We use a threshold value $\Delta_{\rm th} = 0.025$, which is about the average cage size of particles in the glass states at $\varphi_{\rm d}$~\cite{berthier2016growing}: if $\Delta_{\rm r} < \Delta_{\rm th}$, the route to the jammed state belongs to {\it reversible-jamming}; otherwise {to} {\it irreversible-jamming}.

Note that $\{\vec{r}_i^{\rm before} \}$ and  $\{\vec{r}_i^{\rm after} \}$ are both unjammed configurations. We are interested in whether $\{\vec{r}_i^{\rm before} \}$ and  $\{\vec{r}_i^{\rm after} \}$ are 
{close to each other on the free-energy landscape,}
such that they belong to the same glass state.
The reversible-jamming means that, during the route to jamming,  the system remains in the same meta-stable glass basin.
Our definition of reversibility shall be distinguished from the one used in some previous studies~\cite{schreck2013particle, kawasaki2016macroscopic, das2020unified}, where a jammed state is called reversible under cyclic AQS only if all particles return to the exactly same positions at jamming.
{The present definition is essentially the same as the one employed in \cite{jin2018stability} for TQS in thermal HSs.}

\begin{figure}[h]
\centerline{\includegraphics[width=0.7\columnwidth]{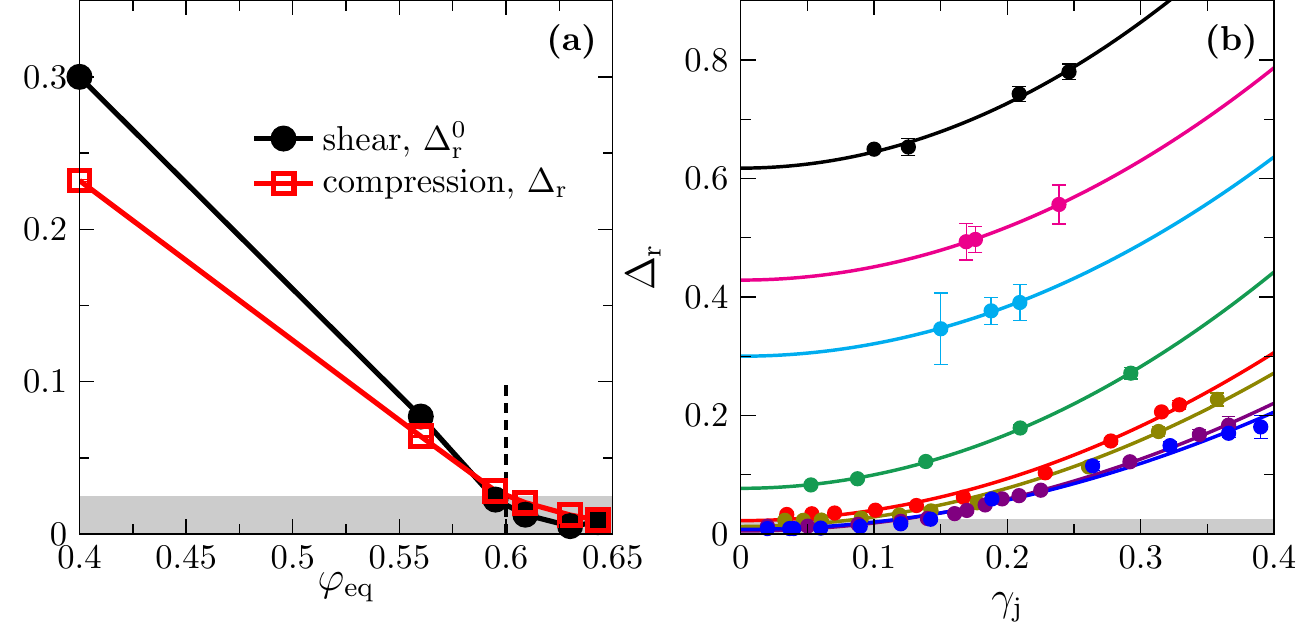} 
  } 
\caption{ 
The RMSD measured by one cycle of  ARC or AQS.
(a) The RMSD $\Delta_{\rm r}$ of an ARC-cycle, and the zero strain RMSD $\Delta_{\rm r}^{0}$ of an AQS-cycle (obtained from (b)), are plotted as functions of $\varphi_{\rm eq}$. 
    The location of $\varphi_{\rm SF} \approx 0.60$ is indicated by the vertical dashed bar.
          The RMSD is smaller than $\Delta_{\rm th}=0.025$ in the shaded region.
  (b) RMSD $\Delta_{\rm r}$ of an AQS-cycle as a function of the jamming strain $\gamma_{\rm j}$, along SJ-lines for $N=8000$ and a few different  $\varphi_{\rm eq}$ (from top to bottom, $\varphi_{\rm eq} = 0, 0.2, 0.4, 0.56, 0.595, 0.609, 0.630, 0.643$). The data points are fitted to an empirical form $\Delta_{\rm r} (\gamma_{\rm j})= \Delta_{\rm r}^0 + c \gamma_{\rm j}^2$ (lines). The zero strain RMSD  $\Delta_{\rm r}^0$ is plotted as a function of $\varphi_{\rm eq}$ in (a).
}
\label{fig:reversibility_quench}
\end{figure}

\subsection{Reversible-jamming and irrversible-jamming}


 First let us study the reversibility of the ARC jamming procedure (\eq{eq:compression_route}).
 In {Sec.}~\ref{sec:athermal}, we have shown that there is a mapping between the isotropic jamming density $\varphi_{\rm IJ}$ and 
 $\varphi_{\rm eq}$, described by 
 Eq.~(1)
 (recall that $\varphi_{\rm IJ}$ is the most probable jamming density $\varphi_{\rm j}$ obtained by ARC for the given  $\varphi_{\rm eq}$, see \eq{eq:fij}).
In Fig.~\ref{fig:reversibility_quench}(a) we show the RMSD
$\Delta_{\rm r}$ measured by one cycle of ARC: $\{\varphi_{\rm eq},0\} \to  \{\varphi_{\rm j}=\varphi_{\rm IJ},0\} \to \{\varphi_{\rm eq},0\}$.
It can be seen that in the range
$\varphi_{\rm eq} > \varphi_{\rm SF} \approx 0.60$,
$\Delta_{\rm r} < \Delta_{\rm th}$
so that  ARC is reversible.
On the other hand, for $\varphi_{\rm eq} < \varphi_{\rm SF}$, $\Delta_{\rm r}$ becomes much larger  $\Delta_{\rm th}$ so that ARC is irreversible.
Interestingly, the point $\varphi_{\rm eq} = \varphi_{\rm SF}$ that separates reversible-jamming and irreversible-jamming coincides with the one that separates 
compression rate-independent (i.e., state-following) and compression rate-dependent quench dynamics (see Fig.~\ref{fig:Jline_HS}). 
We will interpret this observation in {Sec.}~\ref{sec:reversibility-quench}.


Next we analyze the reversibility of AQS \eq{eq:shear_route}, by
considering the fraction $1-f_{\rm IJ}$
of samples that remain unjammed at $\varphi_{\rm j}$ 
after ARC.
We apply one cycle of AQS,  $\{\varphi_{\rm j},0\} \to  \{\varphi_{\rm j},\gamma_{\rm j}(\varphi_{\rm j};\varphi_{\rm eq}) \} \to \{\varphi_{\rm j},0 \}$, and measure the associated RMSD $\Delta_{\rm r}$.
In Fig.~\ref{fig:reversibility_quench}(b) we show how  $\Delta_{\rm r}$ increases
with the jamming strain $\gamma_{\rm j}$ along SJ-lines Eq.~(2).
From the data we  extrapolate the RMSD at the zero strain limit $\Delta_{\rm r}^0 =\Delta_{\rm r}(\gamma_{\rm j} \to 0)$ for each SJ-line,
and plot it as a function of $\varphi_{\rm eq}$ (Fig.~\ref{fig:reversibility_quench}(a)).
The data shows that $\Delta_{\rm r}^0$ grows above $\Delta_{\rm th}$ as $\varphi_{\rm eq}$ decreases below $\varphi_{\rm SF}$, which is consistent with the results obtained by ARC.


In Fig.~2(b),
we show the heat-map of  $\Delta_{\rm r}$ measured by the {AQS cycle}, 
which suggests that the J-plane can be divided into two parts: reversible-jamming and irreversible-jamming. 
Comparing Figs.~2(a) and (b), 
we find that  
the reversible-jamming part  corresponds to the domain that can be accessed by the thermal protocols {(TC/TQS)} with $\varphi_{\rm eq} > \varphi_{\rm SF}$.
Therefore, most of the packings generated by the thermal protocol are reversible, 
while those generated by the athermal protocols {(ARC/AQS)} can be both  reversible and irreversible.
Note that the packings in the reversible-jamming regime are reversible in both thermal and athermal protocols. In fact, other properties of the packings in this regime are also independent of the jamming protocol (see the main text).


Figure~2(b)
shows that  there are two boundary lines between reversible-jamming and irreversible-jamming: the state-following line, which is the thermal SJ-line obtained from $\varphi_{\rm eq} = \varphi_{\rm SF}$,
and the yielding-jamming separation line $\{ \varphi_{\rm c}(\varphi_{\rm eq}), \gamma_{\rm c}(\varphi_{\rm eq}) \}$.
 They are associated to two different mechanisms respectively,  quench dynamics and yielding of HS glasses, as explained in detail below.

\begin{figure}[h]
    \centerline{\includegraphics[width=0.7\columnwidth]{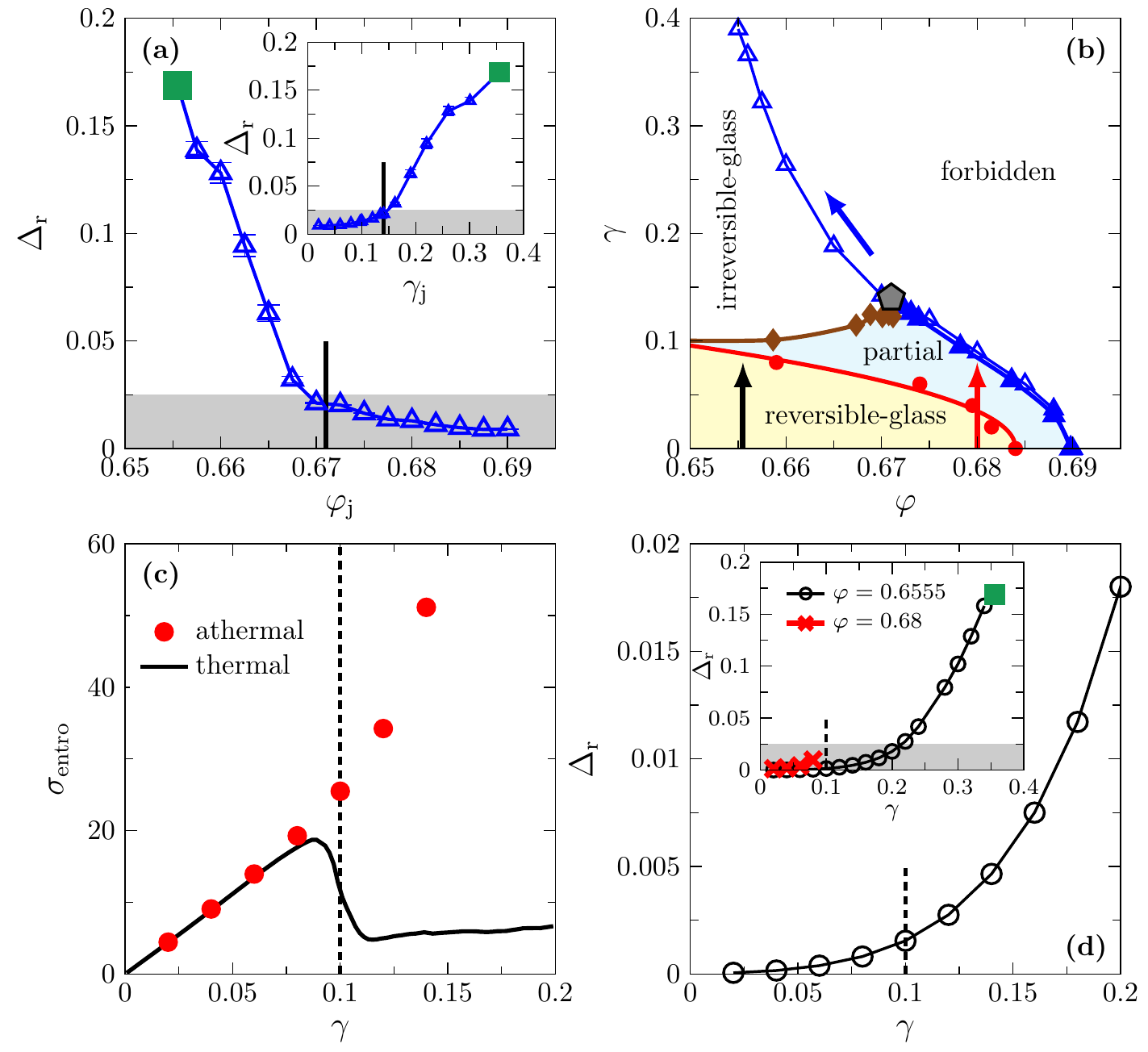}}
\caption{ Connection between irreversible-jamming and HS yielding. Data are obtained for $\varphi_{\rm eq} = 0.643$ and $N=8000$, and the AQS-cycle is used to measure $\Delta_{\rm r}$.
  (a) RMSD $\Delta_{\rm r}$ as a function of $\varphi_{\rm  j}$ and (inset) $\gamma_{\rm j}$ along the SJ-line. The vertical bars represent $\varphi_{\rm c}(\varphi_{\rm eq} = 0.643) = 0.671$, and  (inset) $\gamma_{\rm c}(\varphi_{\rm eq} = 0.643) = 0.14$. 
  The shaded area represents the region with $\Delta_{\rm r} \leq \Delta_{\rm th}$.
  (b) Stability-reversibility map of HS glasses (adapted from Ref.~\cite{jin2018stability}).
The brown diamonds and red circles represent the HS yielding line and the Gardner line. The open and filled triangles are the same SJ-line data as in 
Fig.~2(a).
(c) Entropic stress $\sigma_{\rm entro}$ of unjammed configurations obtained from both athermal and thermal protocols, as functions of $\gamma$, for a fixed  $\varphi = 0.6555 < \varphi_{\rm c}$ (black arrow in (b)). 
(d) RMSD $\Delta_{\rm r}$ as a function of $\gamma$ for $\varphi = 0.6555$. The inset shows the data in a larger range of $\gamma$, as well as the data for $\varphi = 0.68 > \varphi_{\rm c}$ (red arrow in (b)).
The vertical dashed lines in (c) and (d) represent the thermal HS yielding strain {$\gamma_{\rm Y}^{\rm entro}(\varphi = 0.6555, \varphi_{\rm eq} = 0.643)  \approx 0.1$}.
The green squares in (a) and (d) represent the same point at $\{\varphi_{\rm j} = 0.6555, \gamma_{\rm j} = 0.36\}$.
}
\label{fig:yielding}
\end{figure}

\subsection{Connection to quench dynamics}
\label{sec:reversibility-quench}

To explain the above observation, we borrow the framework obtained by a recent mean-field theory of spherical {\it mixed} p-spin model~\cite{folena2020rethinking, zamponi2019surfing} (note the equivalence between the temperature quench in the spin model and the compression quench in our model), {which revealed  some important features missing in usual {\it pure} p-spin models.}

(i) The reversible-jamming regime corresponds to the state-following dynamical regime that only exists for $\varphi_{\rm eq} > \varphi_{\rm SF}$.
A jammed state in this regime is ``followed" from, and only depends on,  the initial equilibrium state at $\varphi_{\rm eq}$. Upon  compression  or shear jamming, the state remains in the same metastable glass basin, and the memory of the initial condition is kept.
Therefore,  the route to jamming is reversible (Figs.~2  
and~\ref{fig:reversibility_quench}), and the function $\varphi_{\rm IJ}(\varphi_{\rm eq})$
is independent of protocol parameters such as the compression rate $\Gamma$ (Fig.~\ref{fig:Jline_HS}).

(ii) The irreversible-jamming regime corresponds to the so-called {\it hic sunt leones} dynamical regime observed in Ref.~\cite{folena2020rethinking}, which exists for $\varphi_{\rm onset} < \varphi_{\rm eq} < \varphi_{\rm SF}$. Such quench dynamics are rather complicated and not fully understood even in the spin glass models~\cite{folena2020rethinking, zamponi2019surfing}. Upon compression  or shear jamming, the memory of the initial condition is partially lost, and the final state is protocol-dependent. 

(iii) The SJ-line, $\gamma_{\rm j} =  \gamma_{\rm j}(\varphi_{\rm j} = \varphi_{\rm J},  \varphi_{\rm eq} < \varphi_{\rm onset})$, which is  vertical in the thermodynamical limit and is the leftmost boundary of the J-plane,  corresponds to the memory-less dynamical regime for $\varphi_{\rm eq} < \varphi_{\rm onset}$.  
The jamming density after quench is always $\varphi_{\rm J}$, which is completely independent of the initial condition.

Based on the above analogy, we attribute  the irreversible-jamming 
 for $\varphi_{\rm eq} < \varphi_{\rm SF}$ to the loss or the partial loss of memory during quench. This mechanism determines one boundary between reversible-jamming and irreversible-jamming, i.e., the state-following line.

\subsection{Connection to yielding of hard sphere glasses}
\label{sec:reversibility-yielding}

In Fig.~\ref{fig:yielding} (a), we plot RMSD $\Delta_{\rm r}$ as a function of $\varphi_{\rm j}$ along the SJ-line Eq.~(3), 
for  $\varphi_{\rm eq} = 0.643$. The reversible-jamming ($\Delta_{\rm r} < \Delta_{\rm th}$) and irreversible-jamming  ($\Delta_{\rm r} > \Delta_{\rm th}$) parts are separated  by $\varphi_{\rm c}(\varphi_{\rm eq} = 0.643) =0.671$, which is also the density separates shear yielding and shear jamming in thermal HSs (see Fig.~\ref{fig:yielding_jamming}  and {Sec.}~\ref{sec:Jplane_th}). 

To understand the reason for the irreversibility when  $\varphi_{\rm j} < \varphi_{\rm c}$,
we analyze how $\Delta_{\rm r}$ increases along the route to shear jamming.
We first use ARC to compress 
the system from $\varphi_{\rm eq}=0.643$ to  $\varphi = 0.6555$ ($\varphi$ is chosen below $\varphi_{\rm c}$), and then apply a cycle of AQS,
$\{\varphi, 0\} \to \{\varphi, \gamma\} \to \{\varphi, 0\}$, at the fixed density $\varphi = 0.6555$.
In Fig.~\ref{fig:yielding} (d) we  plot $\Delta_{\rm r}$ as a function of increasing strain $\gamma$.
Note that, for $\gamma < \gamma_{\rm j}(\varphi = 0.6555, \varphi_{\rm eq}=0.643) = 0.36$, the configuration is not jammed during the cycle of shear. 
The data shows that $\Delta_{\rm r}$ increases with $\gamma$ and becomes significantly larger than $\Delta_{\rm th}$ as $\gamma \to \gamma_{\rm j}$. However, $\Delta_{\rm r}$ is nearly zero
below the  yielding strain {$\gamma_{\rm Y}^{\rm entro}(\varphi = 0.6555, \varphi_{\rm eq} = 0.643) \approx 0.1$} of HS glasses, which suggests that 
the onset of irreversible-jamming might be related to the yielding of HS glasses. 

Considering  that $\Delta_{\rm r}$ is measured in the athermal protocol, the above observation is rather surprising at first glance: why would the behavior of an unjammed athermal system, which is sometimes considered as a ``liquid" (because $P_{\rm mech} = \Sigma_{\rm mech} = 0$)~\cite{o2003jamming},  has anything to do with yielding that typically only occurs in solids?
 To further reveal the connection, we measure  entropic stress-strain curves of unjammed configurations obtained by the two different protocols:
 we
use ARC to compress 
 the system from $\varphi_{\rm eq}=0.643$ to  $\varphi = 0.6555$, shear it using AQS (with {the} SS potential) or TQS (with {the} HS potential)
  up to a strain 
  $\gamma$,  
  and then measure its entropic stress $\sigma_{\rm entro}$ by switching on the temperature (with {the} HS potential).
  Note that as long as $\gamma < \gamma_{\rm j}$, the athermal SS configurations  are unjammed, which allows us to switch to  the HS potential because there is no {overlappings} between particles.
  Figure~\ref{fig:yielding}(c) shows that the entropic stress-strain curves of athermal and thermal configurations coincide below {$\gamma_{\rm Y}^{\rm entro}$}. The bifurcation occurs around {$\gamma_{\rm Y}^{\rm entro}$}: the entropic stress $\sigma_{\rm entro}$ of the athermal system tends to diverges as  $\gamma \to \gamma_{\rm j}$, while that  of the thermal system reaches a plateau after yielding, {$\gamma > \gamma_{\rm Y}^{\rm entro}$}. 
  
 {The above}  surprising  results suggest that, these athermal configurations, even though unjammed, should be understood as glass states rather than liquid states. After yielding, the system leaves the metastable glass basin and explores a larger configurational space~\cite{RUYZ14}. 
  The athermal protocol stops only when it successfully finds a jammed configuration that belongs to a different glass basin, and therefore the route is not reversible. In the thermal protocol, the thermal activations can overcome  free-energy barriers between different glass basins. After yielding, the system eventually reaches a stationary flow  instead of jamming. Note that one can also interpret yielding as the end point of state-following dynamics under shear~\cite{rainone2016following}, which is consistent with the mechanism discussed in {Sec.}~\ref{sec:reversibility-quench}.

We finally make a comparison between the J-plane and the stability-reversibility map of HS glasses obtained in Ref.~\cite{jin2018stability} (Fig.~\ref{fig:yielding}(b)). The HS stability-reversibility map describes stability and reversibility of thermal HS glasses under volume and shear strains. In the {\it reversible-glass} (stable glass) regime, the glass responds elastically 
 to shear. In the {\it partially irreversible-glass} (marginally stable glass) regime, the glass is marginally stable and experiences mesoscopic plastic deformations under shear. At larger strains, the system either yields and becomes irreversible, or jams. 
To avoid confusion, we use reversible-glass/irreversible-glass/partially irreversible-glass for the stability-reversibility map, and reversible-jamming/irreversible-jamming for the J-plane. 
The irreversible-jamming part of the SJ-line lies in the irreversible-glass regime, as a consequence of HS {yielding detected by the measurement
of the entropic stress discussed above.}
The reversible-jamming part of the SJ-line belongs to the partially irreversible-glass regime, because the system remains in the same meta-stable glass state despite of the plasticity (see Fig.~1 of Ref.~\cite{jin2018stability} and the discussion there).

{\section{Stress-controlled athermal quasistatic shear}
\label{sec:stress_controlled}
The initial state is the same as in the strain-controlled AQS (see Sec.~\ref{sec:athermal_protocol}), which is
an unjammed SS system at $\{\varphi < \varphi_{\rm j}(\varphi_{\rm eq}), \gamma=0\}$, obtained from ARC of an equilibrium state at $\{\varphi =\varphi_{\rm eq}, \gamma=0\}$. A target mechanical stress $\Sigma_{\rm mech}^{\rm t}$ is preset. The strain $\gamma$ is increased (if the current stress $\Sigma_{\rm mech} < \Sigma_{\rm mech}^{\rm t}$) or decreased (if $\Sigma_{\rm mech} > \Sigma_{\rm mech}^{\rm t}$) at each step by $\delta \gamma = 10^{-4}$ with energy minimization. The onset of shear jamming at $\gamma \approx \gamma_{\rm j}$ is clearly observed in stress-controlled simulations, at which the stress becomes non-zero and rapidly increases upon further increasing $\gamma$ (Fig.~\ref{fig:stress_controlled}). The values of jamming strain $\gamma_{\rm j}$ obtained by strain-controlled (the star in Fig.~2)
and stress-controlled (Fig.~\ref{fig:stress_controlled}) simulations are \red{basically} consistent; \red{the jamming strain $\gamma_{\rm j}(\varphi_{\rm j}=0.68; \varphi_{\rm eq} = 0.643) \approx 0.08$, shown in Fig.~\ref{fig:stress_controlled}(c), is slightly smaller than  $\gamma_{\rm j}(\varphi_{\rm j}=0.68; \varphi_{\rm eq} = 0.643) \approx 0.09$ in Fig.~2, because smaller strain steps ($\delta \gamma = 10^{-4}$) are used here.
Note that the discrepancy  $\delta \gamma_{\rm j} \approx 0.01$ is within the strain step size $\delta \gamma = 0.02$ (see Sec.~\ref{sec:athermal_protocol}) used to obtain Fig.~2.}
If $\Sigma_{\rm mech}^{\rm t}$ is below the yield stress $\Sigma^{\rm mech}_{\rm Y}$, the strain $\gamma$ does not increase anymore once  $\Sigma_{\rm mech}^{\rm t}$ is reached, and the system remains as a solid. Otherwise, if $\Sigma_{\rm mech}^{\rm t} > \Sigma^{\rm mech}_{\rm Y}$, the strain $\gamma$  will never stop increasing since the system becomes a plastic flow after yielding. Note that the yielding studied here is for athermal SS packings (above jamming), which should not be confused with the yielding in thermal HS glasses (below jamming) discussed in Sec.~\ref{sec:reversibility-yielding}.
}

\begin{figure*}[h]
\centerline{\includegraphics[width=\columnwidth]{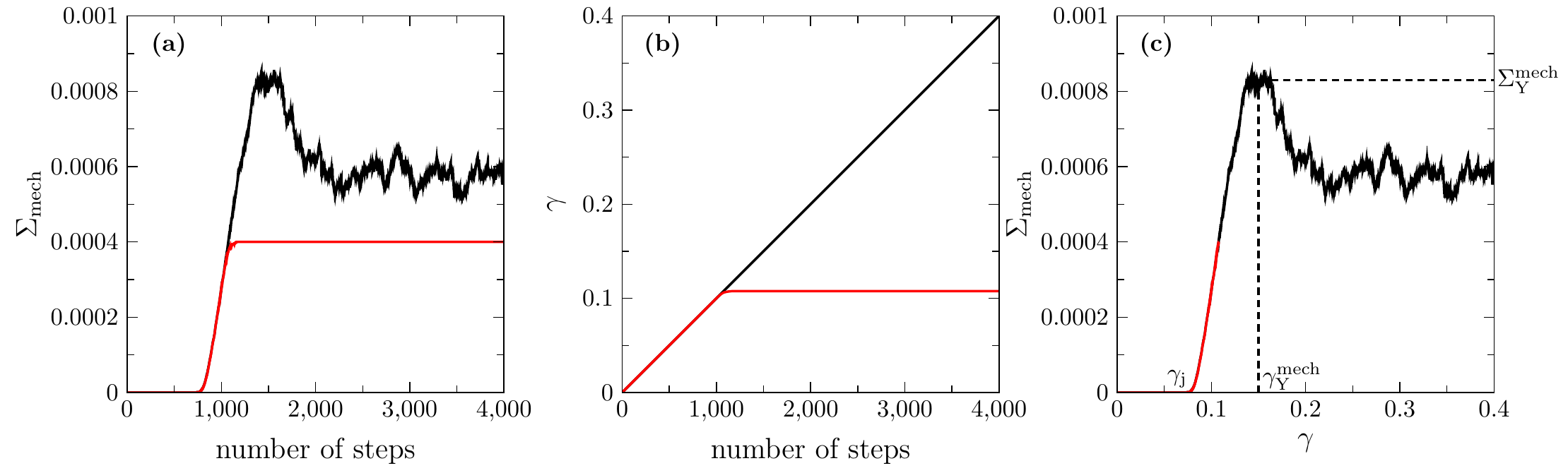}} 
\caption{{Stress-controlled AQS of $N=2000$ SSs at a constant $\varphi = 0.68$ for $\varphi_{\rm eq} = 0.643$. The setup  corresponds to the state point $\{\varphi_{\rm j}=0.68, \gamma_{\rm j}=0.09\}$ in Fig.~2(a)
(star), and according to Ref.~\cite{babu2020friction}, the system's yield stress is $\Sigma^{\rm mech}_{\rm Y} \approx 0.0008$. For two representative cases, $\Sigma_{\rm mech}^{\rm t} = 0.0004 < \Sigma^{\rm mech}_{\rm Y}$ (red curves) and $\Sigma_{\rm mech}^{\rm t} = 0.05 > \Sigma^{\rm mech}_{\rm Y}$ (black curves), we plot (a) the mechanical stress $\Sigma_{\rm mech}$ and (b) the shear strain $\gamma$ as functions of number of steps. The parametric plot $\Sigma_{\rm mech}$ versus $\gamma$ in (c) clearly shows the onset of shear jamming at $\gamma_{\rm j}$, and the yielding at $\gamma = \gamma_{\rm Y}^{\rm mech}$ and  $\Sigma_{\rm mech} = \Sigma^{\rm mech}_{\rm Y}$. The data are averaged over 24 samples.}}
\label{fig:stress_controlled}
\end{figure*}

\section{Shear jamming of face-centered cubic  crystals}
\label{sec:fcc}

In order to understand the differences in shear jamming between amorphous and ordered states, we simulate FCC crystals consisting of $N=500$ particles.  
Similar to the amorphous case, crystalline configurations jam under thermal or athermal shear. 
However, because crystals
are in equilibrium, their states are independent of protocol parameters such as $\varphi_{\rm eq}$ and $\Gamma$.
As a consequence, the J-plane shrinks to a single SJ-line, whose end point is the state point of the FCC close packing at $\{\varphi_{\rm J}^{\rm FCC} \simeq 0.74, \gamma_{\rm j} =0 \}$ (Fig.~\ref{fig:fcc}(a)). 
Near this point, the SJ-line Eq.~(2)
follows a simple {form}, $\left ( 1 - 2 \gamma_{\rm j} \right )^2 + 1 = 4 (D/l_a)^2$, where $l_a$ is the lattice constant of the unit cell, and $D/l_a = (3 \varphi_{\rm j}/ 2\pi)^{1/3}$.
This relationship is derived from
an
affine transformation. 

In the thermal protocol, the FCC crystals also exhibit shear jamming and yielding, which are separated by the density $\varphi_{\rm c} \approx 0.69$ (Fig.~\ref{fig:fcc}(b)). In the shear jamming case ($\varphi > \varphi_{\rm c}$), the stress-strain curve has a small tip before the divergence of the stress $\sigma_{\rm entro}$, which indicates the onset of non-affine arrangement. We interpret the tip as a vestige of yielding. The MRSD $\Delta_{\rm r}$ increases rapidly below $\varphi_{\rm c}$ (Fig.~\ref{fig:fcc}(c)), which suggests that reversible-jamming and irreversible-jamming are separated by $\varphi_{\rm c}$, consistent with the amorphous case.

\begin{figure*}[h]
\centerline{\includegraphics[width=\columnwidth]{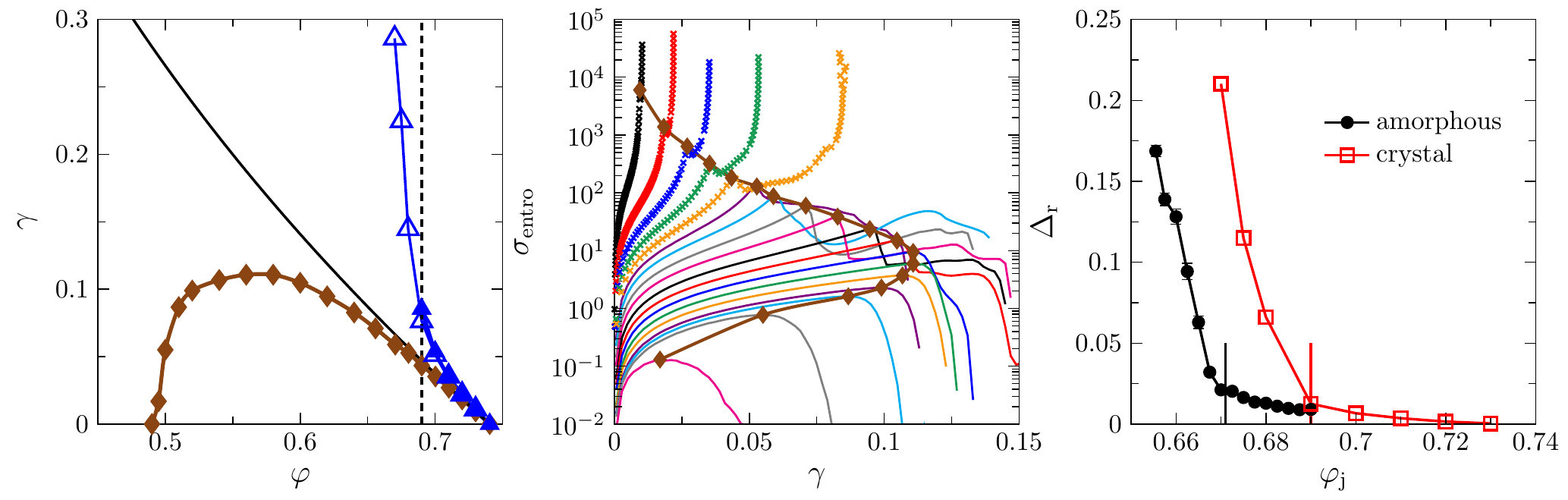}} 
\caption{Shear jamming of FCC crystals ($N=500$). (a) SJ-lines obtained by the thermal (filled triangles) and the athermal (open triangles) protocols (to be compared with Fig.~\ref{fig:yielding}(b)). We also plot the affine law (solid line), the thermal yielding line (diamonds), and $\varphi_{\rm c} \approx 0.69$ (vertical dashed line) estimated from the data in (b).  (b) Stress-strain curves for a few different densities (from bottom to top, $\varphi = 0.495, 0.50, 0.51, 0.52, 0.54, 0.56, 0.58, 0.60, 0.62, 0.64, 0.655, 0.67, 0.68, 0.69, 0.70, 0.71, 0.72, 0.73$). (c) RMSD $\Delta_{\rm r}$ as a function of $\varphi_{\rm j}$ for FCC packings, and for amorphous packings  with $\varphi_{\rm eq} = 0.643$ (same data as in Fig.~\ref{fig:yielding}a). The vertical bars indicate  $\varphi_{\rm c}$. }
\label{fig:fcc}
\end{figure*}

\section{More general jamming protocols}
\label{sec:general_protocols}

\red{So far we have investigated how to obtain jamming using a combination of compression and simple shear. In this section we discuss additional jamming protocols, by considering (i) switching the order of compression and shear, and (ii) more general deformation modes that can involve multiple independent shears.}

\red{\subsection{Switching the order of compression and shear}
We aim to examine whether the SJ-lines (see Eq.~(2)) presented in Fig.~2 depend on the order of compression and shear. Recall that, to obtain these SJ-lines, the system is rapidly compressed  to a target density $\varphi = \varphi_{\rm j}$, and then is sheared up to jamming at a jamming strain $\gamma_{\rm j}$.
As an example, here we consider $N=2000$ systems with $\varphi_{\rm eq} = 0.643$, and use athermal protocols. Figure~\ref{fig:stress_controlled}(c) shows that the onset of shear jamming occurs at $ \gamma_{\rm j} \approx 0.08$, after the system is initially compressed from $\varphi_{\rm eq} = 0.643$ to $\varphi = 0.68$.
To switch the order, we first shear the system to $\gamma = 0.08$ at the fixed density $\varphi = \varphi_{\rm eq} = 0.643$ (with $\delta \gamma = 10^{-4}$), and then apply athermal quasi-static compression (AQC) under the fixed shear strain condition ($\gamma = 0.08$), with step size $\delta \varphi = 5 \times 10^{-4}$. 
 Figure~\ref{fig:shear_comp} shows that the onset of jamming occurs at $ \varphi_{\rm j} \approx 0.68$. As a result, the same  jammed state point $\{\varphi_{\rm j} \approx 0.68, \gamma_{\rm j} \approx 0.08\}$ is obtained irrelevant of the order of compression and shear.
This examination validates that Eqs.~(2) and (3) in the main text can be used equivalently. }
 
\begin{figure*}[ht]
\centerline{\includegraphics[width=0.35\columnwidth]{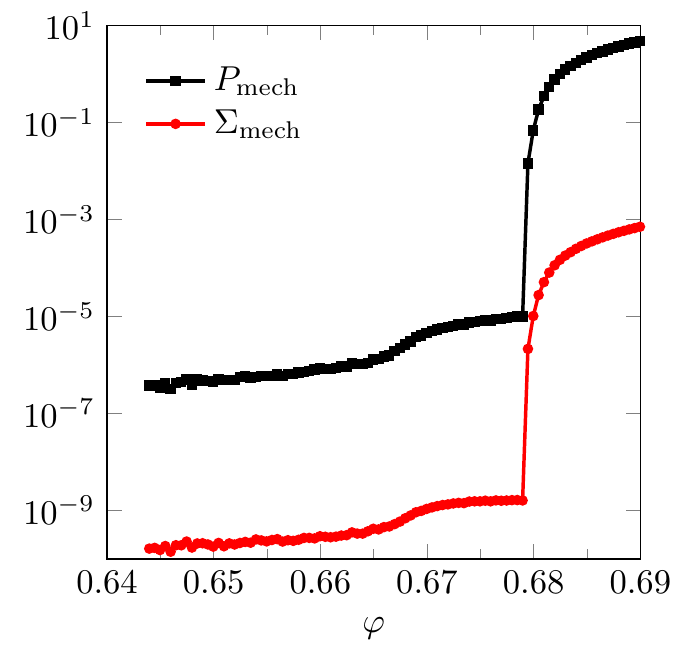}} 
\caption{\red{Pressure-density and stress-density curves of $N=2000$ and $\varphi_{\rm eq} = 0.643$ systems, obtained by AQC at a fixed shear strain $\gamma = 0.08$.  Before AQC, the systems are initially sheared to $\gamma = 0.08$ using AQS, at the fixed density $\varphi = \varphi_{\rm eq} = 0.643$. The onset of jamming at $ \varphi_{\rm j} \approx 0.68$ can be seen from the abrupt increase of both pressure and stress.}}
\label{fig:shear_comp}
\end{figure*}
\begin{figure*}[ht]
\centerline{\includegraphics[width=0.7\columnwidth]{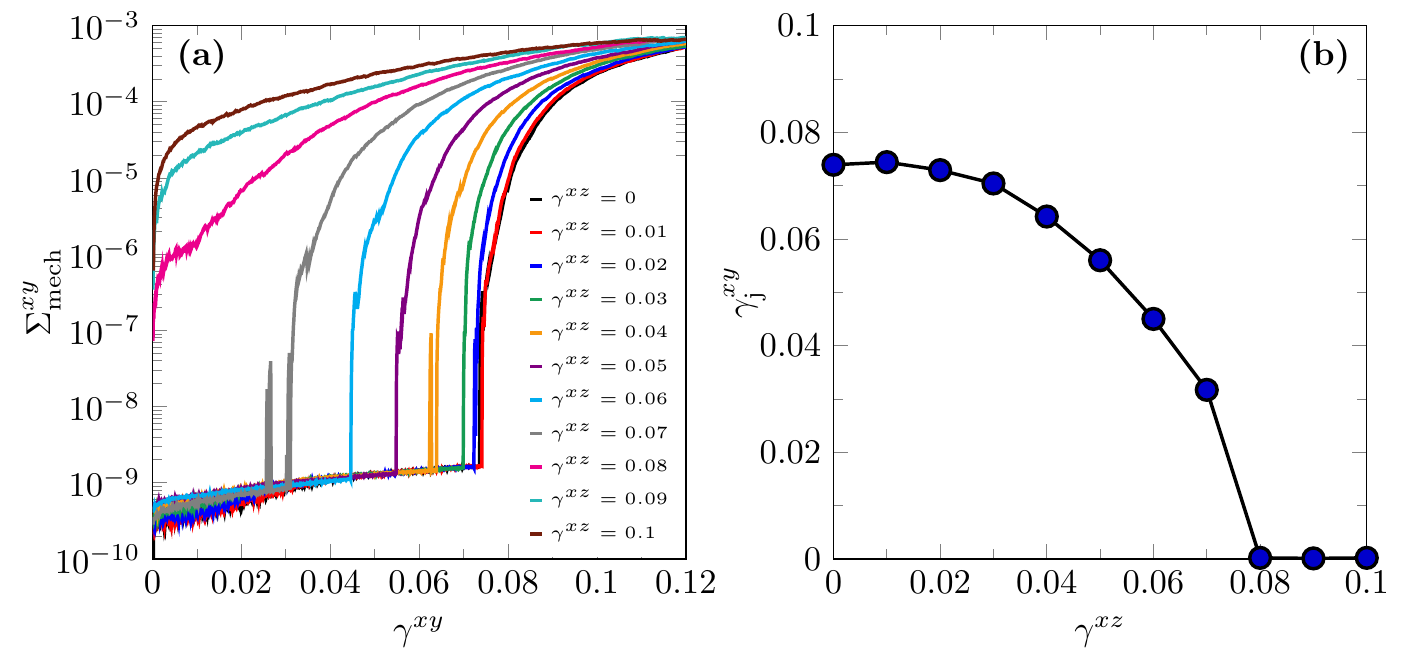}} 
\caption{\red{Jamming by simple shears applied in two perpendicular planes, for $N=2000$ and $\varphi_{\rm eq} = 0.643$ systems, using the AQS protocol. (a) Stress-strain curves obtained by  the simple shear in the $x$-$y$ plane, after applying an initial strain $\gamma^{xz}$  in the $x$-$z$ plane. (b) Jamming strain $\gamma^{xy}_{\rm j}$ as a function of $\gamma^{xz}$.}}
\label{fig:two_shears}
\end{figure*}

\red{\subsection{Isochoric  (fixed-volume) deformations}
We have shown that frictionless spheres can be jammed not only by compression, but also by shear, which conserves the volume.
Here we consider a general isochoric deformation, described by a deformation gradient tensor $\bf{F}$, which makes a linear transformation of an infinitesimal  line element $d \bf{r}$ as, $d {\bf r'} = {\bf  F}  d {\bf r}$. Using the polar decomposition theorem~\cite{brannon2018rotation}, it can be shown that, after rotations, the original deformation can be represented by pure stretches along principal axes, whose deformation gradient tensor has a form,
\beq
{\bf F}  = 
\begin{bmatrix}
\lambda_1 & 0 & 0\\
0 & \lambda_2 & 0\\
0 & 0 & \lambda_3
\end{bmatrix},
\eeq
where the principal stretches obey,
\beq
\lambda_1 \lambda_2 \lambda_3 = 1,
\label{eq:const-volume}
\eeq
in order to satisfy the isochoric condition. Therefore, in general, any  homogeneous isochoric deformation may be reproduced by a combination of  two independent simple (or pure) shears and proper rotations.}

\red{Without loss of generality, we study jamming by two successive  simple shears in perpendicular planes. The system ($\varphi_{\rm eq} = 0.643$ and $N=2000$) is firstly sheared in the $x$-$z$ plane by a strain $\gamma^{xz}$,
and then sheared in the $x$-$y$ plane by a strain $\gamma^{xy}$ (with $\gamma^{xz}$ fixed), using the AQS protocol (step size $\delta \gamma = 10^{-4}$). 
If $\gamma^{xz} < \gamma_{\rm j}^{xz}$, where $ \gamma_{\rm j}^{xz}$ is the shear jamming strain, the system is unjammed after applying the shear strain $\gamma^{xz}$.
In such cases, the onset of shear jamming in the $x$-$y$ direction  is determined from the abrupt increase of stress $\Sigma_{\rm mech}^{xy}$, at  a jamming strain $\gamma_{\rm j}^{xy}$ (see Fig.~\ref{fig:two_shears}(a)). The jamming strain $\gamma_{\rm j}^{xy}$ is negatively correlated to the pre-strain $\gamma^{xz}$ (see Fig.~\ref{fig:two_shears}(b)).
We find that $\gamma_{\rm j}^{xy}(\gamma^{xz} = 0) \approx \gamma_{\rm j}^{xz} \approx 0.08$, which is consistent with the initial isotropic condition,
and the value of jamming strain $\gamma_{\rm j}  \approx 0.08$ obtained by one simple shear (see Fig.~\ref{fig:stress_controlled}(c)). 
}

\red{\subsection{Jamming-space for general deformations} The results in previous subsection suggest that,  for a jamming protocol that is composed by general isochoric deformations and compressions, the J-plane presented in Figs. 1 and 2  will be extended into a three-dimensional  {\it jamming-space} (J-space) on  $\varphi_{\rm j}$-$\gamma^1_{\rm j}$-$\gamma^2_{\rm j}$ axes, where $\gamma^1_{\rm j}$ and $\gamma^2_{\rm j}$ ($\gamma^{xz}_{\rm j}$ and $\gamma^{xy}_{\rm j}$ in the above case) are jamming strains in two orthogonal directions. }

\red{We may consider even more general deformations. Any homogenous deformation can be decomposed into (with rotations) shape changing and  volume changing parts. 
Because the constant-volume constrain Eq.~(\ref{eq:const-volume}) is relaxed,  maximally three independent principle stretches can be defined. Indeed, it is known that any homogeneous strain
 can be produced by the succession of three simple shears in mutually perpendicular planes, a uniform volume change, and a rotation~\cite{truesdell1960classical}. We thus expect a generalized four-dimensional  jamming-space on  $\varphi_{\rm j}$-$\gamma^1_{\rm j}$-$\gamma^2_{\rm j}$-$\gamma^3_{\rm j}$ axes,
 with the jamming density $\varphi_{\rm j}$ and three orthogonal jamming strains $\gamma^1_{\rm j}$, $\gamma^2_{\rm j}$ and  $\gamma^3_{\rm j}$. 
 }

 \bibliography{jamming,jammingSI}

\end{document}